\documentstyle[11pt,th-en-roman]{article}

\newtheorem{Def}{Definition}

\newtheorem{Prop}{Proposition}
\newtheorem{Lem}{Lemma}
\newtheorem{Theo}{Theorem}

\newtheorem{deff}{Definition}[section]

\def\la{\leftarrow}
\def\ra{\rightarrow}
\def\ex{\exists}
\def\fa{\forall}
\def\bd{\noindent\bf}
\def\sbd{\vspace{8pt}\noindent\bf}

\newbox\tempa
\newbox\tempb
\newdimen\tempc
\def\mud#1{\hfil $\displaystyle{\mathstrut #1}$\hfil}
\def\rig#1{\hfil $\displaystyle{#1}$}
\def\irulehelp#1#2#3{\setbox\tempa=\hbox{$\displaystyle{\mathstrut #2}$}%
		        \setbox\tempb=\vbox{\halign{##\cr
	\mud{#1}\cr
	\noalign{\vskip\the\lineskip}%
	\noalign{\hrule height 0pt}%
	\rig{\vbox to 0pt{\vss\hbox to 0pt{${\; #3}$\hss}\vss}}\cr
	\noalign{\hrule}%
	\noalign{\vskip\the\lineskip}%
	\mud{\copy\tempa}\cr}}%
		      \tempc=\wd\tempb
		      \advance\tempc by \wd\tempa
		      \divide\tempc by 2 }
\def\irule#1#2#3{{\irulehelp{#1}{#2}{#3}%
		     \hbox to \wd\tempa{\hss \box\tempb \hss}}}

\begin{document}
\begin{center}
{\Large{\bf A Complete Proof Synthesis Method for the Cube of Type Systems}}
\\[10pt]
{\bf Gilles Dowek}\\[10pt]
{\bf INRIA}
\def\thefootnote{\fnsymbol{footnote}}
\footnote[1]{B.P. 105, 78153 Le Chesnay CEDEX, France. Gilles.Dowek@inria.fr}
\footnote[2]{This research was partly supported by ESPRIT Basic Research Action
``Logical Frameworks''.}
\def\thefootnote{\arabic{footnote}}
\\[15pt]
\end{center}

\section*{Abstract}

We present a complete proof synthesis method for the eight type systems of
Barendregt's cube extended with $\eta$-conversion. Because these systems
verify the proofs-as-objects paradigm, the proof synthesis method is a one 
level process merging unification and resolution.
Then we present a variant of this method, which is incomplete but much more 
efficient. At last we show how to turn this algorithm into a unification 
algorithm.

\section*{Introduction}

The Calculus of Constructions is an extension of Church higher order logic in 
which the terms belong to a $\lambda$-calculus with dependent types, 
polymorphism and type constructors.
Because of the richness of the terms structure, proofs can be represented as 
terms through Heyting semantics and Curry-Howard isomorphism. So a proof of a 
proposition $P$ is merely a term of type $P$.

A proof synthesis method for Church higher order logic is given in
\cite{Huet72}. In this paper we generalize this method to find terms of a given
type in the systems of Barendregt's cube (in particular in the Calculus of
Constructions) extended with $\eta$-conversion.
In some sense, because these type systems are more powerful than Church
higher order logic, the proof synthesis problem is more complicated.
But we claim that because the proofs-as-objects paradigm simplifies the 
formalisms, it also simplifies the proof synthesis algorithms. In particular 
resolution and unification can be merged in a uniform algorithm.

We first present a complete method, then we discuss some efficiency 
improvements, we present a variant of this method which is incomplete but much
more efficient and we show how to turn this algorithm into a unification 
algorithm.

\section{Outline of the Method}

\subsection{Resolution and Unification}

In first order logic or higher order logic, we have two syntactical categories,
terms and proofs. Terms are trees (in first order logic) or simply 
typed $\lambda$-terms (in higher order logic) and proof are build, for 
instance, with natural deduction rules. These rules combine some proofs and 
terms to form new proofs. For instance the rule $\ra$-elim
$$\irule{A \ra B~~~~A} {B} {}$$
combines two proofs (one of $A \ra B$ and one of $A$) to get a proof of $B$.
The rule $\fa$-elim
$$\irule{\fa x:A.B~~~~t:A} {B[x \la t]} {}$$
combines a proof (of $\fa x:A.B$) and a term (of type $A$) to get a proof of
$B[x \la t]$. So proofs are heterogeneous trees. For instance, in the proof
$$\irule{\irule{\fa f:T \ra T.((P~f) \ra (Q~f))~~~~
\irule{[x:T]}{[x:T]x:T \ra T}{}}
{(P~[x:T]x) \ra (Q~[x:T]x)}{}
~~~~~~~~~~~~~~(P~[x:T]x)}
{(Q~[x:T]x)}{}$$
the subtree
$$\irule{[x:T]}{[x:T]x:T \ra T}{}$$
is a term derivation.

In first order logic and higher order logic, the {\it resolution} algorithm 
searches for proof-trees. When during the search of a proof-tree, a term-tree 
is needed, the {\it unification} algorithm is called.
In type systems, term-trees and proof-trees belong to a single syntactical
category, so our algorithm is a one level process merging resolution and 
unification.

We shortly present now the unification and resolution algorithms
for higher order logic \cite{Huet72} \cite{Huet75} \cite{Huet76}, our goal is 
to show that a single idea on term enumeration underlies both algorithms.

\subsection{Higher Order Unification}

Higher order unification is based on an algorithm that enumerates all the 
normal $\eta$-long terms of a given type in the simply typed 
$\lambda$-calculus.

Let $P_{1} \ra ... \ra P_{n} \ra P$ ($P$ atomic) be a type.
All normal $\eta$-long terms of this type begin by $n$ abstractions
$$t = [x_{1}:P_{1}] ... [x_{n}:P_{n}] t'$$
The term $t'$ must have type $P$ so it is an atomic term. 
A variable $w:Q_{1} \ra ... \ra Q_{p} \ra Q$ of the context or which is an 
$x_{i}$ can be the head of this term, only if $Q = P$. Then
this variable must be applied $p$ times in order to get a term of the good 
type.
So
$$t = [x_{1}:P_{1}] ... [x_{n}:P_{n}](w~c_{1}~...~c_{p})$$
To mark the dependency of the $c_{i}$ on the $x_{j}$ we write
$$t = [x_{1}:P_{1}] ... [x_{n}:P_{n}]
(w~(d_{1}~x_{1}~...~x_{n})~...~(d_{p}~x_{1}~...~x_{n}))$$
To find the $d_{i}$ we use recursively the same algorithm. So we first take
$$t = [x_{1}:P_{1}] ... [x_{n}:P_{n}]
(w~(h_{1}~x_{1}~...~x_{n})~...~(h_{p}~x_{1}~...~x_{n}))$$
then we use recursively our algorithm to instantiate the
$h_{i}:P_{1} \ra ... \ra P_{n} \ra Q_{i}$.

In unification, flexible-rigid equations are solved by instantiating the 
head of the flexible term using this method,
the choice of the variable $w$ is very restricted by the rigid term.
Rigid-rigid equations are simplified. And flexible-flexible 
equations are solved in a trivial way.

So the method to enumerate terms underlying higher order unification can be
summaried
\begin{itemize}
\item
try all the possible head variables of the normal $\eta$-long form of the term,
\item
generate new variables for the rest of the term,
\item
generate a typing constraint to enforce well-typedness of the term,
\item
use recursively this method to fill these variables.
\end{itemize}

\subsection{Higher Order Resolution}

Let us present now the resolution method. 
We just present an incomplete restriction of this method to 
Horn clauses i.e. propositions on the form 
$Q_{1} \ra ... \ra Q_{n} \ra Q$, 
where $Q_{1}, ..., Q_{n}, Q$ are atomic propositions (free variables 
are considered as universally quantified in the head of
the clause).

When the goal $P$ is atomic we unify it with $Q$ the head of an hypothesis
$Q_{1} \ra ... \ra Q_{n} \ra Q$,
we generate subgoals $\sigma Q_{1}, ..., \sigma Q_{n}$ (where $\sigma$ is a
unifer of $P$ and $Q$) and we recursively search proofs of these propositions.

When the goal is also a Horn clause $P_{1} \ra ... \ra P_{n} \ra P$,
then the clause form of the negation of this goal introduces
$P_{1}$, ..., $P_{n}$ as hypothesis and the atomic goal $P$.

This restriction is called the {\it introduction-resolution} algorithm.
It can be presented with two rules: the resolution rule between an 
atomic goal and the head of an hypothesis is just called the {\it resolution} 
rule and the introduction of the hypothesis $P_{1}$ when we want to prove 
$P_{1} \ra P$ is called the {\it introduction} rule.
It is incomplete. In \cite{Huet72} another rule (the {\it splitting} rule) is
added to make it complete.

Remark that this method can also be used for {\it hereditary Horn clauses}
i.e. propositions on the form 
$Q_{1} \ra ... \ra Q_{n} \ra Q$,
where $Q$ is atomic and $Q_{1}, ..., Q_{n}$ are hereditary Horn clauses.

\subsection{Introduction-Resolution in Type Systems}

In type systems, all the propositions can be considered as hereditary Horn 
clauses where the arrow is generalized to a dependent product, so we can apply 
the same method.

Furthermore, in type systems, a proof-synthesis method must not only assert
that the proposition is provable but also exhibit a proof-term of this 
proposition.

The introduction-resolution algorithm can be presented that way:
\begin{itemize}
\item
{\it Introduction}: To find a proof of a proposition $(x:P_{1})P$
in a context $\Gamma$, find a proof of $P$ in the context $\Gamma[x:P_{1}]$.
When we have a proof $t$ of $P$, the term $[x:P_{1}]t$ is a proof of 
$(x:P_{1})P$ in the context $\Gamma$.
\item {\it Resolution}: To find a proof of $P$ in a context $\Gamma$ where
there is a proposition
$$f:(x_{1}:Q_{1}) ... (x_{n}:Q_{n})Q$$
unify $P$ and $Q$ (consider $x_{1}, ..., x_{n}$ as variables in 
$Q$) and then if $x_{i}$ has not been instantiated during the unification
process, find a proof of $\sigma Q_{i}$ (where $\sigma$ is a unifier of
$P$ and $Q$). 
For each $i$, we get (by unification or as a proof of a subgoal) a term $c_{i}$
of type $Q_{i}[x_{1} \la c_{1}, ..., x_{i-1} \la c_{i-1}]$. 
The term $(f~c_{1}~...~c_{n})$ is a proof of $P$.
\end{itemize}
This method is fully described in \cite{Helmink}.
Let us give an example. We have an hypothesis
$$f:(x:T)(y:T)(z:T)(u:(R~x~y))(v:(R~y~z))(R~x~z)$$
and we want to prove $(R~a~c)$. We unify $(R~x~z)$ and $(R~a~c)$ this
gives the substitution $x \la a, z \la c$. Then we have to find terms $b:T$,
$t:(R~a~b)$ and $u:(R~b~c)$.
The term $(f~a~b~c~t~u)$ is a proof of $(R~a~c)$.

This method is in general incomplete. Another drawback is that no general 
unification 
algorithm is known for type systems. Algorithms are known only for the simply 
typed $\lambda$-calculus \cite{Huet75} \cite{Huet76} and 
the $\lambda \Pi$-calculus \cite{Elliott89} \cite{Elliott90} \cite{Pym}.

An alternative presentation of this method is the following
\begin{itemize}
\item
{\it Introduction}: we give the proof $[x:P_{1}]h$ for the proposition
and the variable $h$ is a subgoal to be proved.
\item
{\it Resolution}: we give the proof $(f~h_{1}~...~h_{p})$ for the proposition 
together with an equation $Q[x_{1} \la h_{1}, ..., x_{p} \la h_{p}] = P$
(in our example $(R~h_{1}~h_{3}) = (R~a~b)$) to force the term
$(f~h_{1}~...~h_{p})$ to have type $P$.
We solve this unification problem which fills some of the variables 
$h_{i}$ (here $h_{1}$ and $h_{3}$) and then the other variables 
(here $h_{2}$, $h_{4}$ and $h_{5}$) are subgoals to be proved.
\end{itemize}
When, to prove a proposition, we first apply the introduction rule $n$ times 
then the resolution rule, we give the proof 
$[x_{1}:P_{1}] ... [x_{n}:P_{n}](f~h_{1}~...~h_{p})$ i.e. we 
give $f$ for the head variable of proof.
Remark that the dependence of the $h_{i}$ on the $x_{j}$ is implicit.

So the method to enumerate terms underlying the introduction-resolution 
method is
\begin{itemize}
\item
try all the possible head variables of the normal $\eta$-long form of the term,
\item
generate new variables for the rest of the term,
\item
generate a typing constraint to enforce well-typedness of the term,
\item
use recursively this method to fill these variables.
\end{itemize}
So this method is the same as the method underlying higher order unification.

\subsection{A Method to Enumerate the Terms of a Given Type}

Now we use this idea to construct a complete proof synthesis method for type 
systems. To enumerate the terms $t$ that we can substitute to a variable $x$
of type $T$ we imagine the normal $\eta$-long form of $t$
$$t = [x_{1}:P_{1}]...[x_{n}:P_{n}](w~c_{1}~...~c_{p})$$
or
$$t = [x_{1}:P_{1}]...[x_{n}:P_{n}](y:A)B$$
and we perform all the elementary substitutions
$$x \la [x_{1}:P_{1}]...[x_{n}:P_{n}]
(w~(h_{1}~x_{1}~...~x_{n})~...~(h_{p}~x_{1}~...~x_{n}))$$
$$x \la [x_{1}:P_{1}]...[x_{n}:P_{n}](y:(h_{1}~x_{1}~...~x_{n}))
(h_{2}~x_{1}~...~x_{n}~y)$$
Then we enumerate all the terms that can be substituted to the variables 
$h_{1}, ..., h_{p}$.

Because we consider terms in $\eta$-long form, 
the number of abstractions $n$ 
is the number of products in the type $T$ and the types $P_{1}, ..., P_{n}$ 
are the types of the variables bound in these products. So for completeness we
have to consider
\begin{itemize}
\item 
all the possible $w$,
\item
all the possible $p$ and types for $h_{1}, ..., h_{p}$.
\end{itemize}
When we perform such a substitution, we have to make sure to get a term of
type $T$. This well-typedness constraint is an equation.

In the introduction-resolution algorithm, the solution of this constraint is 
called the unification step of resolution. In unification in simply typed
$\lambda$-calculus, types are always ground, so the constraint relates
ground terms an we only have to check that they are identical, this is 
the selection of the head variables step. In unification in the
$\lambda \Pi$-calculus \cite{Elliott89} \cite{Elliott90} \cite{Pym} this
constraint is added to the set of equations, it is the {\it accounting}
equation.

Here we keep this equation in the context. Equations are constraints that
the forthcoming substitutions must verify.
Since we do not have a special algorithm to solve these equations but use the 
standard method to fill the variables and use the equations as constraints 
on the substitutions,  this method is a one level process merging resolution 
and unification.

\subsection{Number of Applications}

In the introduction-resolution algorithm when we want to prove an atomic 
proposition $P$ and we resolve it with a variable
$w:(y_{1}:Q_{1}) ... (y_{q}:Q_{q})Q$ ($Q$ atomic) we only consider proofs of 
the form $(w~h_{1}~...~h_{q})$ and the constraint is that the type of this 
term must be equal to $P$ i.e. $Q[y_{1} \la h_{1}, ..., y_{q} \la h_{q}] = P$.

But in polymorphic type systems the term $(w~c_{1}~...~c_{q})$ can still
have a type which is a product and we can go on applying variables.
This is why the introduction-resolution method is incomplete and why
a splitting rule is needed in \cite{Huet72}.

Here we consider an infinite number of 
possibilities for the number of applications and an infinitely branching 
search tree, we use interleaving to enumerate its nodes. In an incomplete
but more efficient method we consider a restriction which is more or less
similar to the introduction-resolution method.

\subsection{Variables in the Proposition to be Proved}

In some type systems, the variables
$h_{i}$ may have occurrences in the type of $h_{j}$ for $j>i$. We could
first instantiate the variable $h_{i}$ and then instantiate the variable
$h_{j}$ when it has a ground type, but it is well known that it is more
efficient to solve the variable $h_{j}$ first. For instance to enumerate the 
terms of the type $T = \ex x:Nat.(P~x)$ we must not enumerate all the terms 
$n$ of type $Nat$ and for each $n$ enumerate the terms of type $(P~n)$, but 
enumerate first the terms of type $(P~x)$.

Thus in polymorphic type systems when we have a variable in the proposition
to be proved, it is not always possible to know the number and
the types of abstractions: if the proposition to be proved is
$T = (x_{1}:P_{1}) ... (x_{n}:P_{n})P$ ($P$ atomic)
and the head of $P$ is a variable that can be substituted then a substitution 
may increase the number of products in $T$ and therefore the number of 
abstractions in $t$.
So we have to delay the instantiation of $x:T$ until we have instantiated 
$T$ in a term $(x_{1}:P_{1}) ... (x_{n}:P_{n})P$ where the head of $P$ cannot
be substituted. 

\subsection{$\eta$-conversion}

In this paper we consider pure type systems extended with 
$\eta$-conversion. The methods developed here should also be applicable
for pure type systems without $\eta$-conversion but, as in higher order 
unification \cite{Huet75} \cite{Huet76}, we would need to consider more 
elementary substitutions. 

\section{Pure Type Systems Extended with $\eta$-conversion}

\subsection{Definition}

{\it A pure type system extended with $\eta$-conversion} \cite{Barendregt}
is a $\lambda$-calculus given by a set $S$ (the elements of $S$ are called 
{\it sorts}), a subset $Ax$ of $S \times S$ and a subset $R$ of 
$S \times S \times S$.

\begin{Def} Functional Type System

A type system is said to be {\it functional} if
$$<s,s'> \in Ax~\mbox{and}~<s,s''> \in Ax~\mbox{implies}~s' = s''$$
$$<s,s',s''> \in R~\mbox{and}~<s,s',s'''> \in R~\mbox{implies}~s'' = s'''$$
\end{Def}

In this paper we consider systems such that $S = \{Prop,Type\}$, 
$Ax = \{<Prop,Type>\}$,
if $<s,s',s''> \in R$ then $s' = s''$ and $<Prop,Prop,Prop> \in R$.
These systems are the eight systems of Barendregt's cube. All these systems 
are functional. Examples are the simply typed
$\lambda$-calculus, the $\lambda \Pi$-calculus \cite{Howard} \cite{deBruijn80}
\cite{HHP}, the system F, the system F$\omega$ \cite{Girard} and the Calculus 
of Constructions \cite{Coquand85} \cite{CoqHue88}.

\begin{Def} Syntax

$$T~::=~s~|~x~|~(T~T)~|~[x:T]T~|~(x:T)T$$
\end{Def}

In this paper we ignore variable renaming problems. A rigorous presentation
would use de Bruijn indices \cite{deBruijn72}. The terms $s$ are sorts, the 
terms $x$ are called variables, the terms $(T~T')$ applications, the terms
$[x:T]T'$ $\lambda$-abstractions and the terms $(x:T)T'$ products. 
The notation $T \ra T'$ is used for $(x:T)T'$ when $x$ has no free occurrence 
in $T'$.

Let $t$ and $t'$ be terms and $x$ a variable. We write $t[x \la t']$ for the 
term obtained by substituting $t'$ for $x$ in $t$. We write 
$t \equiv t'$ when $t$ and $t'$ are $\beta \eta$-equivalent.

\begin{Def} Context

A {\it context} $\Gamma$ is a list of pairs $<x,T>$ (written $x:T$) where $x$ 
is a variable and $T$ a term. The term $T$ is called the {\it type} of $x$ in 
$\Gamma$.

We write $[x_{1}:T_{1}; ...; x_{n}:T_{n}]$ for the context with elements 
$x_{1}:T_{1}, ..., x_{n}:T_{n}$ and $\Gamma_{1}\Gamma_{2}$ for the 
concatenation of the contexts $\Gamma_{1}$ and $\Gamma_{2}$.
\end{Def}

\begin{Def} Typing Rules

We define inductively two judgements: {\it $\Gamma$ is well-formed} and 
{\it $t$ has type $T$ in $\Gamma$} ($\Gamma \vdash t:T$) where $\Gamma$ is
a context and $t$ and $T$ are terms.\\
Empty:
$$\irule{} 
        {[~]~\mbox{well-formed}}
        {}$$
Declaration:
$$\irule{\Gamma \vdash T:s} 
        {\Gamma[x:T]~\mbox{well-formed}}
        {s \in S}$$
Sort:
$$\irule{\Gamma~\mbox{well-formed}} 
        {\Gamma \vdash s:s'}
        {<s,s'> \in Ax}$$
Variable:
$$\irule{\Gamma~\mbox{well-formed}~~x:T \in \Gamma} 
        {\Gamma \vdash x:T}
        {}$$
Product:
$$\irule{\Gamma \vdash T:s~~\Gamma[x:T] \vdash T':s'}
        {\Gamma \vdash (x:T)T':s''}
        {<s,s',s''> \in R}$$
Abstraction:
$$\irule{\Gamma \vdash (x:T)T':s~~\Gamma[x:T] \vdash t:T'} 
        {\Gamma \vdash [x:T]t:(x:T)T'}
        {s \in S}$$
Application:
$$\irule{\Gamma \vdash t:(x:T)T'~~\Gamma \vdash t':T}
        {\Gamma \vdash (t~t'):T'[x \la t']}
        {}$$
Conversion: 
$$\irule{\Gamma \vdash T:s~~\Gamma \vdash T':s~~\Gamma \vdash t:T~~T \equiv T'}
        {\Gamma \vdash t:T'}
        {s \in S}$$
\end{Def}

\begin{Def} Well-typed Term

A term $t$ is said to be {\it well-typed} in a context $\Gamma$ if there exists
a term $T$ such that $\Gamma \vdash t:T$.
\end{Def}

\begin{Def} Type

A term $T$ is said to be a {\it type} in the context $\Gamma$ if it is a sort
or if there exists a sort $s$ such that $\Gamma \vdash T:s$.
\end{Def}

\begin{Prop} If $\Gamma \vdash t:T$ then $T$ is a type.

{\bd Proof} By induction on the length of the derivation of 
$\Gamma \vdash t:T$.
\end{Prop}

\begin{Prop}
The $\beta \eta$-reduction is strongly normalizable and confluent 
on well-typed terms. Thus a well-typed term has a unique $\beta \eta$-normal 
form. Two well-typed terms are equivalent if they have the same 
$\beta \eta$-normal form.

{\bd Proof} Proofs of strong normalization and confluence for $\beta$-reduction
are given in \cite{Coquand85} \cite{GeuNed}. As remarqued in 
\cite{Geuvers} the strong normalization proof of \cite{GeuNed} can be adapted 
to $\beta \eta$-reduction and does not need confluence. Then confluence
proofs for $\beta \eta$-reduction are given in \cite{Geuvers} \cite{Salvesen}
\cite{Coquand91}.
\end{Prop}

\begin{Prop} In a functional type system, in which reduction is confluent, 
a term $t$ well-typed in a context $\Gamma$ has a unique type modulo 
$\beta \eta$-equivalence.

{\bd Proof} By induction over the structure of $t$.
\end{Prop}

\begin{Def} Atomic Term

A term $t$ is said to be {\it atomic} if it has the form $(u~c_{1}~...~c_{n})$
where $u$ is a variable or a sort. The symbol $u$ is called the {\it head}
of the term $t$.
\end{Def}

\begin{Prop}
A normal well-typed term $t$ is either an abstraction, a product or an atomic 
term.

{\bd Proof}
If the term $t$ is neither an abstraction nor a product then it can be written 
in a unique way 
$$t = (u~c_{1}~...~c_{n})$$ 
where $u$ is not an application. 
The term $u$ is not a product (if $n \neq 0$ because a product is of type 
$s$ for some sort $s$ and therefore cannot be applied and if $n = 0$ because
$t$ is not a product).
It is not an abstraction (if $n \neq 0$ because $t$ is in normal form
and if $n = 0$ because $t$ is not an abstraction). It 
is therefore a variable or a sort.
\end{Prop}

\begin{Prop}
Let $T$ be a well-typed normal type, the term $T$ can 
be written in a unique way $T = (x_{1}:P_{1}) ... (x_{n}:P_{n})P$ with $P$
atomic.

{\bd Proof} By induction over the structure of $T$.
\end{Prop}

\begin{Def} $\eta$-long form

Let $\Gamma$ be a context and $t$ be a $\beta \eta$-normal term well-typed in 
$\Gamma$ and $T$ the $\beta \eta$-normal form of its type.
The {\it $\eta$-long form} of the term $t$ is defined as
\begin{itemize}
\item
If $t = [x:U]u$ then the $\eta$-long form of $t$ is $[x:U']u'$ where $U'$ is 
the $\eta$-long form of $U$ in $\Gamma$ and $u'$ the $\eta$-long form of $u$ 
in $\Gamma[x:U]$.
\item
If $t = (x:U)V$ then the $\eta$-long form of $t$ is $(x:U')V'$ where $U'$ is 
the $\eta$-long form of $U$ in $\Gamma$ and $V'$ the $\eta$-long form of $V$ 
in $\Gamma[x:U]$.
\item If $t = (w~c_{1}~...~c_{p})$ then we let 
$T = (x_{1}:P_{1})...(x_{n}:P_{n})P$ ($P$ atomic). The $\eta$-long form of $t$
is $[x_{1}:P'_{1}] ... [x_{n}:P'_{n}](w~c'_{1}~...~c'_{p}~x'_{1}~...~x'_{n})$ 
where $c'_{i}$ is the $\eta$-long form of $c_{i}$ in $\Gamma$, $P'_{i}$ the 
$\eta$-long form of $P_{i}$ in $\Gamma[x_{1}:P_{1}; ...; x_{i-1}:P_{i-1}]$
and $x'_{i}$ the $\eta$-long form of $x_{i}$ in
$\Gamma[x_{1}:P_{1}; ...; x_{i}:P_{i}]$.
\end{itemize}
The well-foundedness of this definition is proved in \cite{matching} 
\cite{These}.
\end{Def}

\begin{Def} normal $\eta$-long form

Let $t$ be a term well-typed in a context $\Gamma$, the {\it normal 
$\eta$-long form} of $t$ is the $\eta$-long form of its $\beta \eta$-normal 
form.
\end{Def}

\begin{Def} Subterm

We consider well-typed normal $\eta$-long terms labeled with the contexts in 
which they are well-typed: $t_{\Gamma}$.
Let $t_{\Gamma}$ such a term, we define by induction over the structure of 
$t_{\Gamma}$ the set $Sub(t_{\Gamma})$ of {\it strict subterms} of 
$t_{\Gamma}$
\begin{itemize}
\item
if $t_{\Gamma}$ is a sort or a variable then $Sub(t_{\Gamma}) = \{\}$,
\item
if $t_{\Gamma}$ is an application, $t = (u~v)$, then
$Sub(t_{\Gamma}) = \{u_{\Gamma}, v_{\Gamma}\} \cup Sub(u_{\Gamma}) \cup 
Sub(v_{\Gamma})$,
\item
if $t_{\Gamma}$ is an abstraction, $t = [x:P]u$, then
$Sub(t_{\Gamma}) = \{P_{\Gamma},u_{\Gamma[x:P]}\} \cup Sub(P_{\Gamma}) \cup 
Sub(u_{\Gamma[x:P]})$,
\item
if $t_{\Gamma}$ is a product, $t = (x:P)u$, then
$Sub(t_{\Gamma}) = \{P_{\Gamma},u_{\Gamma[x:P]}\} \cup Sub(P_{\Gamma}) \cup
Sub(u_{\Gamma[x:P]})$.
\end{itemize}
\end{Def}

\subsection{The $Meta$ Type System}

Let ${\cal T}$ be a type system of the cube.
In order to express the proof synthesis method, we have to be allowed
to declare a variable that stands for any well-typed term of ${\cal T}$.
For instance a variable that stands for $Prop$ i.e. a variable of type
$Type$ and this is not possible in ${\cal T}$ because the term $Type$ is 
not well-typed.

We want also to be allowed to 
express a term $t:T'$ which is well-typed in $\Gamma[x:T]$ as $t = (f~x)$ with
$f$ well-formed in $\Gamma$. We want actually to be allowed to define 
$f = [x:T]t$ whatever the terms $T$ and $t$ may be. 
In general this cannot be done in ${\cal T}$, because the type $(x:T)T'$ may 
be not well-typed. So we are going to embed our type system ${\cal T}$ in 
another type system: the {\it $Meta$} type system.

\begin{Def} $Meta$
$$Meta = <S',Ax',R'>$$ 
$$S' = \{Prop,Type,Extern\}$$ 
$$Ax' = \{<Prop,Type>,<Type,Extern>\}$$
$$R' = \{<Prop,Prop,Prop>,<Prop,Type,Type>,<Type,Prop,Prop>,$$
$$<Type,Type,Type>,<Prop,Extern,Extern>,<Type,Extern,Extern>\}$$
\end{Def}

The strong normalization and confluence of $\beta \eta$-reduction for 
this system are not yet fully proved. But since all the terms typable in the 
$Meta$ type system are
typable in the Calculus of Constructions with Universes \cite{Coquand86}
\cite{Luo} (identifying the sorts $Extern$ and $Type_{1}$), the 
$\beta$-reduction is strongly normalizable and confluent 
\cite{Coquand86} \cite{Luo}.
It is conjectured in \cite{Geuvers} that in a pure type system in which the
$\beta$-reduction is strongly normalizable the 
$\beta \eta$-reduction also is. Then assuming strong normalization, 
confluence is proved in \cite{Geuvers} \cite{Salvesen} \cite{Coquand91}.

\begin{Prop} Because the $Meta$ type system is functional and 
$\beta \eta$-reduction is confluent, if a term $t$ is well-typed in a
context $\Gamma$ then it has a unique type modulo $\beta \eta$-equivalence.
\end{Prop}

\section{Constrained Quantified Contexts and Substitution}

\subsection{Constrained Quantified Contexts}

\begin{Def} Constrained Quantified Contexts 

A {\it quantified declaration} is a triple $<Q,x,T>$ (written $Qx:T$) 
where $Q$ is a quantifier ($\fa$ or $\ex$), $x$ a variable and $T$ is a
term. A {\it constraint} is a pair of terms $<a,b>$ (written $a = b$).
A {\it constrained quantified context} is a list of quantified declarations and
constraints. If $\Gamma$ contains the declaration
$\fa x:T$ then the variable $x$ is said to be {\it universal} in $\Gamma$.
If it contains the declaration $\ex x:T$ then $x$ is said to be 
{\it existential} in $\Gamma$. These constrained 
quantified contexts are generalizations of Miller's {\it mixed prefixes} 
\cite{Miller} which are lists of quantified declarations.
\end{Def}

Usual contexts are identified with constrained quantified contexts with only 
universal variables.

\begin{Def} Equivalence Modulo Constraints

Let $\Gamma$ be a constrained quantified context, we
define the relation between terms $\equiv_{\Gamma}$ as the smallest 
equivalence relation compatible with terms structure such that
\begin{itemize}
\item
if $t \equiv t'$ then $t \equiv_{\Gamma} t'$,
\item
if $(a = b) \in \Gamma$ then $a \equiv_{\Gamma} b$.
\end{itemize}
\end{Def}

\begin{Def} Typing Rules

First we modify the rules to deal with the new syntax.\\
The {\it declaration} rule is modified in
$$\irule{\Gamma \vdash T:s} 
        {\Gamma[Qx:T]~\mbox{well-formed}}
        {s \in S}$$
the {\it variable} rule is modified in
$$\irule{\Gamma~\mbox{well-formed}~~Qx:T \in \Gamma} 
        {\Gamma \vdash x:T}
        {}$$
the {\it product} rule is modified in
$$\irule{\Gamma \vdash T:s~~\Gamma[Qx:T] \vdash T':s'}
        {\Gamma \vdash (x:T)T':s''}
        {<s,s',s''> \in R}$$
the {\it abstraction} rule is modified in
$$\irule{\Gamma \vdash (x:T)T':s~~\Gamma[Qx:T] \vdash t:T'} 
        {\Gamma \vdash [x:T]t:(x:T)T'}
        {s \in S}$$
and we add a {\it constraint} rule
$$\irule{\Gamma \vdash a:T~~\Gamma \vdash b:T} 
        {\Gamma[a = b]~\mbox{well-formed}}
        {}$$
Then we extend the system by replacing the {\it conversion} rule by
$$\irule{\Gamma \vdash T:s~~\Gamma \vdash T':s~~\Gamma \vdash t:T~~
         T \equiv_{\Gamma} T'}
        {\Gamma \vdash t:T'}
        {s \in S}$$
This defines two new judgements: $\Gamma$ is {\it well-formed using the
constraints} and {\it $t$ has type $T$ in $\Gamma$
using the constraints}.
\end{Def}

Remark that a term may be well-typed in $\Gamma$ using the constraints and
still be not normalizable.

\begin{Def} Well-typed Without Using the Constraints

Let $\Gamma$ be a context and $t$ and $T$ be two terms. The term $t$ is said 
to be {\it of type $T$ in $\Gamma$ without using the constraints} 
if there exists $\Delta$ subcontext of $\Gamma$ (i.e. obtained by removing
some items of $\Gamma$) such that $\Delta$ has no
constraints, is a well-formed context and $\Delta \vdash t:T$.
\end{Def}

\begin{Prop} If a term is well-typed in a context without using the
constraints then it is strongly normalizable.
\end{Prop}

\begin{Def} Normal Form of a Context

Let $\Gamma$ be a well-formed context, the {\it normal form} of $\Gamma$ is 
obtained by 
normalizing (in normal $\eta$-long form) all the types of variables that are 
well-typed without using the constraints and all the constraints which terms 
are well-typed without using the constraints.
\end{Def}

\begin{Prop}
Let $\Gamma$ be a well-formed context and $\Gamma'$ be its normal form. 
The context $\Gamma'$ is well-formed and if $\Gamma \vdash t:T$ then 
$\Gamma' \vdash t:T$.

{\bd Proof}
By induction on the length of the derivation of {\it $\Gamma$ well-formed} and
$\Gamma \vdash t:T$.
\end{Prop}

\begin{Def} Ground Term

A term $t$ is said to be {\it ground} in a context $\Gamma$ if it has no 
occurrence of an existential variable.
\end{Def}

\begin{Def} Rigid and Flexible Terms

A normal $\eta$-long term well-typed without the constraints 
which is an abstraction, a product or an 
atomic term $(w~c_{1}~...~c_{n})$ with $w$ universal variable or sort is said 
to be {\it rigid}.
A normal $\eta$-long term well-typed without the constraints
which is atomic $(w~c_{1}~...~c_{n})$ with $w$ existential variable is said
to be {\it flexible}.
\end{Def}

\begin{Def} Success and Failure Contexts

A normal well-formed context $\Gamma$ is said to be 
{\it a success context} if it has only universal variables and constraints
relating identical terms.
It is said to be {\it a failure context} if it contains a constraint relating 
two normal $\eta$-long ground terms which are not identical.
\end{Def}

\begin{Prop}
Let $\Gamma$ be a normal context which is neither
a success context nor a failure context. Then there exists an existential
variable $x:T$ whose type $T$ is well-typed without using the constraints and 
has the form
$(x_{1}:P_{1}) ... (x_{n}:P_{n})P$ with the term $P$ atomic and rigid in the 
context $\Gamma  [\fa x_{1}:P_{1}; ...; \fa x_{n}:P_{n}]$.

{\bd Proof}
Let us consider the leftmost item which is neither a universal variable
nor a normal $\eta$-long constraint relating identical terms (such an item 
exists since the context is not a success context).

If this item is a constraint $(a = b)$, we let 
$\Gamma = \Delta[a = b]\Delta'$, the terms $a$ and $b$ are well-typed in 
$\Delta$ so, since there are neither existential variables nor (non trivial) 
constraints in $\Delta$, 
they are ground, well-typed without the constraints and normal. These 
terms are different and the context $\Gamma$ is a failure context. So 
if $\Gamma$ is not a failure context then this item is an existential 
variable $\ex x:T$ and $T$ is well-typed without the constraints and ground.
\end{Prop}

\subsection{Substitution}

\begin{Def} Existential Context

A context (well-formed or not) is said to be {\it existential} if it 
contains only declarations of existential variables and constraints but no
declarations of universal variables.
\end{Def}

\begin{Def} Substitution

A finite set $\sigma$ of triples $<x,\gamma,t>$ where $x$ is a variable, 
$\gamma$ an existential context and $t$ a term is said to be a substitution if
for each variable $x$ there is at most one triple of the form $<x,\gamma,t>$ 
in $\sigma$. When $\sigma$ contains only one triple $<x,\gamma,t>$ we write it 
$\sigma = x \la \gamma,t$.
\end{Def}

\begin{Def} Variable Bound by a Substitution

Let $x$ be a variable and $\sigma$ be a substitution. If the substitution 
$\sigma$ contains a triple $<x,\gamma,t>$ then $x$ is said to be {\it bound} 
by $\sigma$. The context $\gamma$ is said to be {\it the context associated to
$x$ in $\sigma$}.
\end{Def}

\begin{Def} Substitution Applied to a Term

Let $x$ be a variable and $\sigma$ a substitution. If there is a triple 
$<x,\gamma,t>$ in $\sigma$ then we let $\sigma x = t$ else we let 
$\sigma x = x$. This definition extends straightforwardly to terms by
\begin{itemize}
\item 
$\sigma s = s$,
\item
$\sigma (t~u) = (\sigma t~\sigma u)$,
\item
$\sigma [x:T]u = [x:\sigma T]\sigma u$,
\item
$\sigma (x:T)u = (x:\sigma T)\sigma u$.
\end{itemize}
\end{Def}

\begin{Def} Substitution Applied to a Context

Let $\Gamma$ be a context and $\sigma$ a substitution, the context 
$\sigma \Gamma$ is defined inductively by
\begin{itemize}
\item 
$\sigma [~] = [~]$
\item 
$\sigma (\Delta [Q x:T]) = (\sigma \Delta) \gamma$ 
where $\gamma$ is the context associated to $x$ by $\sigma$ if the variable 
$x$ is bound by $\sigma$ and $\gamma = [Q x:\sigma T]$ otherwise.
\item 
$\sigma (\Delta [a = b]) = (\sigma \Delta)[\sigma a = \sigma b]$
\end{itemize}
\end{Def}

\begin{Def} Substitution Well-typed in a Context

The substitution $\sigma$ is said to be {\it well-typed} in a context
$\Gamma$ if and only if the context $\sigma \Gamma$ is well-formed,
no universal variable of $\Gamma$ is bound by $\sigma$ and 
for each existential variable $x$ of type $T$ in $\Gamma$ and bound by 
$\sigma$, we have 
$(\sigma \Delta) \gamma \vdash t:\sigma T$ where $\Delta$ is the 
unique context such that $\Gamma = \Delta [\ex x:T] \Delta'$ and 
$<x,\gamma,t>$ the unique triple of $\sigma$ binding the variable $x$. 
\end{Def}

\begin{Prop}
Let $\Gamma$ be a context, $\sigma$ a substitution and $T$ and $T'$ terms 
such that $T \equiv_{\Gamma} T'$. 
Then $\sigma T \equiv_{\sigma \Gamma} \sigma T'$.

{\bd Proof} By induction on the length of the derivation of 
$T \equiv_{\Gamma} T'$.
\end{Prop}

\begin{Prop} Let $\sigma$ be a substitution, $t$ and $u$ two terms and
$x$ a variable which is not bound by $\sigma$ and which is not free in any
$\sigma y$ for $y \neq x$. Then 
$\sigma (t[x \la u]) = (\sigma t)[x \la \sigma u]$.

{\bd Proof} By induction over the structure of $t$. 
\end{Prop}

\begin{Prop} Let $\Gamma$ be a context, $\sigma$ a substitution
well-typed in $\Gamma$ and $x$ a variable declared of type $T$ in $\Gamma$. 
Then $\sigma \Gamma \vdash \sigma x:\sigma T$.

{\bd Proof} Since the substitution $\sigma$ is well-typed in $\Gamma$, the 
context $\sigma \Gamma$ is well-formed. If $x$ not bound by $\Gamma$
then $\sigma x = x$ and $Q x:\sigma T \in \sigma \Gamma$ so 
$\sigma \Gamma \vdash \sigma x:\sigma T$.
If $x$ is bound by $\sigma$ then there exists a unique triple 
$<x,\gamma,t>$ in $\sigma$. Let us write $\Gamma = \Delta [\ex x:T] \Delta'$.
We have $\sigma x = t$. Since $\sigma$ is well-typed in $\Gamma$ we have 
$(\sigma \Delta) \gamma \vdash t:\sigma T$ and $(\sigma \Delta) \gamma$ is a 
prefix of $\sigma \Gamma$. So $\sigma \Gamma \vdash \sigma x:\sigma T$.
\end{Prop}

\begin{Prop} 
Let $\Gamma$ be a context, $\sigma$ a substitution well-typed in $\Gamma$,
$t$ and $T$ two terms such that $\Gamma \vdash t:T$. We have
$\sigma \Gamma \vdash \sigma t:\sigma T$.

{\bd Proof} By induction on the length of the derivation of 
$\Gamma \vdash t:T$.
\begin{itemize}
\item
  If the last rule is the {\it sort} rule then we use the fact that 
  $\sigma \Gamma$ is well-formed.
\item
  If the last rule is the {\it variable} rule then we use the fact that
  $\sigma \Gamma$ is well-formed and 
  $\sigma \Gamma \vdash \sigma x:\sigma T$.
\item
  If the last rule is the {\it product} rule then by induction hypothesis we
  have $\sigma \Gamma \vdash \sigma T:s$ and 
  $(\sigma \Gamma)[\fa x:\sigma T] \vdash \sigma T':s'$, so
  $\sigma \Gamma \vdash \sigma (x:T)T':s''$.
\item
  If the last rule is the {\it abstraction} rule then by induction 
  hypothesis we have $\sigma \Gamma \vdash \sigma (x:T)T':s$ and  
  $(\sigma \Gamma)[\fa x:\sigma T] \vdash \sigma t:\sigma T'$, so 
  $\sigma \Gamma \vdash \sigma [x:T]t:\sigma (x:T)T'$.
\item
  If the last rule is the {\it application} rule then by induction
  hypothesis we have $\sigma \Gamma \vdash \sigma t:\sigma (x:T)T'$ and
  $\sigma \Gamma \vdash \sigma t':\sigma T$, so we get
  $\sigma \Gamma \vdash \sigma (t~t'):(\sigma T')[x \la \sigma t']$
  and since $x$ is not bound by $\sigma$ and is not free in any $\sigma y$ 
  for $y \neq x$, we have
  $\sigma \Gamma \vdash \sigma (t~t'):\sigma (T'[x \la t'])$.
\item
  If the last rule is the {\it conversion} rule then by induction 
  hypothesis we have $\sigma \Gamma \vdash \sigma T:s$, 
  $\sigma \Gamma \vdash \sigma T':s$ and since $T \equiv_{\Gamma} T'$
  we have $\sigma T \equiv_{\sigma \Gamma} \sigma T'$, so
  $\sigma \Gamma \vdash \sigma t:\sigma T'$.
\end{itemize}
\end{Prop}

\begin{Def} Composition of Substitutions

Let $\sigma$ and $\tau$ two substitutions. The substitution 
$\tau \circ \sigma$ is defined as
$$(\tau \circ \sigma) = \{<x, \tau \gamma,\tau t>~|~<x,\gamma,t> \in \sigma\} 
\cup \{<x,\gamma,t>~|~<x,\gamma,t> \in \tau~\mbox{and $x$ not bound by 
$\sigma$}\}$$
\end{Def}

\begin{Prop}
Let $\sigma$ and $\tau$ two substitutions and $t$ be a term, we have 
$(\tau \circ \sigma) t = \tau \sigma t$.

{\bd Proof} By induction over the structure of $t$.
\end{Prop}

\begin{Prop}
Let $\sigma$ and $\tau$ two substitutions and $\Gamma$ be a context, we have 
$(\tau \circ \sigma) \Gamma = \tau \sigma \Gamma$.

{\bd Proof} By induction over the length of $\Gamma$.
\end{Prop}

\begin{Prop} 
Let $\Gamma$ be a context and $\sigma$ and $\tau$ two substitutions, such that
$\sigma$ is well-typed in $\Gamma$ and $\tau$ is well-typed in $\sigma \Gamma$
then $\tau \circ \sigma$ is well-typed in $\Gamma$.

{\bd Proof} We have $(\tau \circ \sigma) \Gamma = \tau \sigma \Gamma$, so 
the context $(\tau \circ \sigma) \Gamma$ is well-formed. No universal
variable of $\Gamma$ is bound by $\tau \circ \sigma$. 
If $\Gamma = \Delta [\ex x:T] \Delta'$ then 
$\sigma  (\Delta [\ex x:T]) \vdash \sigma x:\sigma T$ so using the previous
proposition 
$\tau \sigma  (\Delta [\ex x:T]) \vdash \tau \sigma x:\tau \sigma T$, i.e.
$(\tau \circ \sigma) (\Delta [\ex x:T]) \vdash 
(\tau \circ \sigma) x:(\tau \circ \sigma) T$.
\end{Prop}

\section{A Complete Method}

\subsection*{Informal Introduction}

When we search a proof of a proposition $T$ in a context $\Gamma$
we search a substitution $\sigma$ well-typed in $\Gamma[\ex x:T]$ 
such that $\sigma (\Gamma[\ex x:T])$ is a success context. 
When we have a context which contains several existential variables we 
choose such a variable $x$ of type $T = (x_{1}:P_{1})...(x_{n}:P_{n})P$ with
$P$ atomic rigid in $\Gamma[\fa x_{1}:P_{1}; ...; \fa x_{n}:P_{n}]$ and we 
perform an elementary substitution instantiating this variable.
In the general case the elementary substitutions have the form
$$x \la [x_{1}:P_{1}]...[x_{n}:P_{n}]
(w~(h_{1}~x_{1}~...~x_{n})~...~(h_{p}~x_{1}~...~x_{n}))$$
Let us write $(y_{1}:Q_{1})(y_{2}:(Q_{2}~y_{1})) ... 
(y_{q}:(Q_{q}~y_{1}~...~y_{q-1}))(Q~y_{1}~...~y_{q})$ the type of $w$. 
For the substitutions such that $p = q$ we only need to declare the new
variables $h_{1}, ..., h_{q}$ used in the substitution and the constraint 
expressing the well-typedness of the substitution
$$(\vec{x}:\vec{P})(Q~(h_{1}~x_{1}~...~x_{n})~...~(h_{q}~x_{1}~...~x_{n})) =
(\vec{x}:\vec{P})P$$
(where $(\vec{x}:\vec{P})t$ is an abbreviation for 
$(x_{1}:P_{1}) ... (x_{n}:P_{n})t$). 
For the substitutions such that $p = q + r$, $r \geq 1$, we first declare the
variables $h_{1}, ..., h_{q}$, then we need the type of the term 
$(w~(h_{1}~x_{1}~...~x_{n})~...~(h_{q}~x_{1}~...~x_{n}))$ to be a product,
we introduce therefore two existential variables $H_{1}$ and $K_{1}$ and a 
constraint
$$(\vec{x}:\vec{P})(Q~(h_{1}~x_{1}~...~x_{n})~...~(h_{q}~x_{1}~...~x_{n})) = 
(\vec{x}:\vec{P})(z:(H_{1}~x_{1}~...~x_{n}))(K_{1}~x_{1}~...~x_{n}~z)$$
Then we can introduce the variable $h_{q+1}$ with the type
$(\vec{x}:\vec{P})(H_{1}~x_{1}~...~x_{n})$. 
If $r = 1$ we just need a last constraint expressing the well-typedness of the
substitution
$$(\vec{x}:\vec{P})(K_{1}~x_{1}~...~x_{n}~(h_{q+1}~x_{1}~...~x_{n})) = 
(\vec{x}:\vec{P})P$$
and in the general case we introduce in the same way $4 r$ items 
and then a well-typedness constraint.

We also have to include other elementary substitutions in which the 
variable $x$ is substituted by a term of the form 
$[x_{1}:P_{1}]...[x_{n}:P_{n}]u$ where $u$ is a sort or a product.

\begin{Def} Elementary Substitutions 

Let $\Gamma$ be a normal context well-formed in the $Meta$ type system 
such that the type of the type of universal variables is $Prop$ or $Type$ but 
not $Extern$ and which is neither a success nor a failure context.
We choose in $\Gamma$ an existential variable $x$ whose type is normal 
$\eta$-long and has the form 
$T = (x_{1}:P_{1})...(x_{n}:P_{n})P$ with $P$ atomic rigid in 
$\Gamma[\fa x_{1}:P_{1}; ...; \fa x_{n}:P_{n}]$ (such a variable exists
because the context in neither a success nor a failure context) and 
we construct the set of substitutions $\Sigma(\Gamma)$ in the following way.

{\sbd Notation} 
If $t$ is a term, $(\vec{x}:\vec{P})t$ is an abbreviation for 
$(x_{1}:P_{1}) ... (x_{n}:P_{n})t$.
\begin{itemize}
\item
For every $w$ which is a universal variable declared in the left of $x$ 
or which is an $x_{i}$
$$\Gamma[\fa x_{1}:P_{1}; ...; \fa x_{n}:P_{n}] \vdash
w:(y_{1}:Q'_{1}) ... (y_{q}:Q'_{q})Q'~\mbox{($Q'$ atomic)}$$
We let
$$Q_{1} = Q'_{1}$$
$$Q_{2} = [y_{1}:Q'_{1}]Q'_{2}$$
$$...$$
$$Q_{q} = [y_{1}:Q'_{1}] ... [y_{q-1}:Q'_{q-1}]Q'_{q}$$
$$Q = [y_{1}:Q'_{1}] ... [y_{q}:Q'_{q}]Q'$$
We have
$$w:(y_{1}:Q_{1})(y_{2}:(Q_{2}~y_{1})) ... (y_{q}:(Q_{q}~y_{1}~...~y_{q-1}))
(Q~y_{1}~...~y_{q})$$
For every $r \geq 0$ we consider the following {\it substitutions with 
$r$-splitting}. 
Let $s$ be the sort type of $Q'$ and $s'$ the sort type of $P$ in 
$\Gamma[\fa x_{1}:P_{1}; ...; \fa x_{n}:P_{n}]$.
For every sequence $s_{1}, s'_{1}, ..., s_{i}, s'_{i}, ..., s_{r}, s'_{r}$
such that $<s_{1},s'_{1},s> \in R$, ..., $<s_{i},s'_{i},s'_{i-1}> \in R$
and $s'_{r} = s'$, we consider the substitution 
$$x \la \gamma,t$$
where 
$$t = [x_{1}:P_{1}]...[x_{n}:P_{n}]
(w~(h_{1}~x_{1}~...~x_{n})~...~(h_{q+r}~x_{1}~...~x_{n}))$$ 
$$\gamma = \varphi\chi_{1} ... \chi_{r}\psi$$
with
$$\varphi = [\ex h_{1}:(\vec{x}:\vec{P})Q_{1};$$
$$\ex h_{2}:(\vec{x}:\vec{P})(Q_{2}~(h_{1}~x_{1}~...~x_{n}));$$
$$...;$$
$$\ex h_{q}:(\vec{x}:\vec{P})(Q_{q}~(h_{1}~x_{1}~...~x_{n})~...~
(h_{q-1}~x_{1}~...~x_{n}))]$$
$$\chi_{1} = [\ex H_{1}:(\vec{x}:\vec{P})s_{1};$$
$$\ex K_{1}:(\vec{x}:\vec{P})(z:(H_{1}~x_{1}~...~x_{n})) s'_{1};$$
$$(\vec{x}:\vec{P})
(Q~(h_{1}~x_{1}~...~x_{n})~...~(h_{q}~x_{1}~...~x_{n}))
= (\vec{x}:\vec{P})
(z:(H_{1}~x_{1}~...~x_{n}))(K_{1}~x_{1}~...~x_{n}~z);$$
$$\ex h_{q+1}:(\vec{x}:\vec{P})(H_{1}~x_{1}~...~x_{n})]$$
for all $i$, $1 < i \leq r$
$$\chi_{i} = [\ex H_{i}:(\vec{x}:\vec{P})s_{i};$$
$$\ex K_{i}:(\vec{x}:\vec{P})(z:(H_{i}~x_{1}~...~x_{n})) s'_{i};$$
$$(\vec{x}:\vec{P})(K_{i-1}~x_{1}~...~x_{n}~(h_{q+i-1}~x_{1}~...~x_{n}))
= (\vec{x}:\vec{P})(z:(H_{i}~x_{1}~...~x_{n}))(K_{i}~x_{1}~...~x_{n}~z);$$
$$\ex h_{q+i}:(\vec{x}:\vec{P})(H_{i}~x_{1}~...~x_{n})]$$
and if $r = 0$
$$\psi = [(\vec{x}:\vec{P})
(Q~(h_{1}~x_{1}~...~x_{n})~...~(h_{q}~x_{1}~...~x_{n})) =
(\vec{x}:\vec{P})P]$$
otherwise
$$\psi =
[(\vec{x}:\vec{P})(K_{r}~x_{1}~...~x_{n}~(h_{q+r}~x_{1}~...~x_{n})) = 
(\vec{x}:\vec{P})P]$$
\item
If $P$ is a sort then for every sort $s$, such that $<s,P> \in Ax$, we consider
also the substitution 
$$x \la [~],[x_{1}:P_{1}]...[x_{n}:P_{n}]s$$
and for every pair of sorts $<s,s'>$ such that $<s,s',P> \in R$ 
we also consider the substitution
$$x \la \gamma,t$$
where
$$t = 
[x_{1}:P_{1}]...[x_{n}:P_{n}](y:(h~x_{1}~...~x_{n}))(k~x_{1}~...~x_{n}~y)$$
$$\gamma = [\ex h:(\vec{x}:\vec{P})s~;
 ~\ex k:(\vec{x}:\vec{P})(y:(h~x_{1}~...~x_{n}))s']$$
\end{itemize}
The set $\Sigma(\Gamma)$ is the set which contains all the substitutions
considered above. 
\end{Def}

\begin{Def} Derivation

A {\it derivation} of a context $\Gamma$ is a list of substitutions 
$[\sigma_{1}; ...; \sigma_{m}]$ such that 
$\sigma_{i} \in \Sigma(\sigma_{i-1} \sigma_{i-2} ... \sigma_{1} \Gamma)$ 
and $\sigma_{m} \sigma_{m-1} ... \sigma_{1} \Gamma$ is a success context.
\end{Def}

\begin{Def} Search tree

Let $\Gamma$ be a context, we build a tree, called {\it the search tree} of 
$\Gamma$. Nodes are labeled by contexts and edges by elementary substitutions.
The root is labeled by $\Gamma$. Nodes labeled by success and failure 
contexts are leaves. From a node labeled by a context $\Delta$ which is neither
a success context nor failure context, for each $\sigma$ of 
$\Sigma(\Delta)$ we grow an edge labeled $\sigma$ to a new node labeled by
$\sigma \Delta$.

Success nodes are in bijection with the derivations of $\Gamma$.
A semi-algorithm of proof synthesis is to enumerate the nodes of the 
tree in order to find a success node.
Since the number of sons of a node may be infinite we have, in order to get 
an finitely branching tree, to delay the exploration of node with 
$r$-splitting for $r$ generations.

\end{Def}

\section{Examples}

\subsection{A First Order Example}

Let $\Delta = [\fa T:Prop; \fa R:T \ra T \ra Prop; \fa Eq:T \ra T \ra Prop;$\\
$\fa Antisym:(x:T)(y:T)((R~x~y) \ra (R~y~x) \ra (Eq~x~y));
\fa a:T; \fa b:T; \fa u:(R~a~b); \fa v:(R~b~a)]$\\
and $\Gamma = \Delta [\ex x:(Eq~a~b)]$.\\
For $x$, we perform the substitution with head $Antisym$ and $0$-splitting
$$x \la (Antisym~h_{1}~h_{2}~h_{3}~h_{4})$$
We get the context
$\Delta  [\ex h_{1}:T; \ex h_{2}:T; \ex h_{3}:(R~h_{1}~h_{2}); 
\ex h_{4}:(R~h_{2}~h_{1});(Eq~h_{1}~h_{2}) = (Eq~a~b)]$.\\
For $h_{1}$, we perform the substitution with head $a$ and $0$-splitting
$$h_{1} \la a$$
For $h_{2}$, we perform the substitution with head $b$ and $0$-splitting
$$h_{2} \la b$$
We get the context $\Delta  [\ex h_{3}:(R~a~b); \ex h_{4}:(R~b~a)]$
(we do not write the trivial constraints).\\
For $h_{3}$, we perform the substitution with head $u$ and $0$-splitting
$$h_{3} \la u$$
We get the context $\Delta  [\ex h_{4}:(R~b~a)]$.\\
For $h_{4}$, we perform the substitution with head $v$ and $0$-splitting
$$h_{4} \la v$$
And we get the success context $\Delta$.\\
Let $\theta$ be the composition of these substitutions, 
$\theta x = (Antisym~a~b~u~v)$.

As we will see in the following, in this example, at each step, all the 
substitutions but the one considered leed to obviously hopeless contexts.

\subsection{An Example with Splitting}

Let $\Delta = [\fa A:Prop; \fa B:Prop; \fa I:Prop \ra Prop; 
\fa u:(P:Prop)((I~P) \ra P);$\\
$\fa v:(I~(A \ra B)); \fa w:A]$\\
and $\Gamma = \Delta [\ex p:B]$.\\
For $p$, we perform the substitution with head $u$ and $1$-splitting
$$p \la (u~h_{1}~h_{2}~h_{3})$$
with the new variables:
$h_{1}:Prop$, $h_{2}:(I~h_{1})$, $H:Prop$, $K:H \ra Prop$ and $h_{3}:H$.\\
We get the context\\
$\Delta[\ex h_{1}:Prop; \ex h_{2}:(I~h_{1});\ex H:Prop; 
\ex K:H \ra Prop; \ex h_{3}:H; h_{1} = (x:H)(K~x); (K~h_{3}) = B]$.\\
For $h_{1}$, we perform a product substitution
$$h_{1} \la (x:h')(k'~x)$$
We get the context\\ 
$\Delta[\ex h':Prop; \ex k':h' \ra Prop; \ex h_{2}:(I~((x:h')(k'~x))); 
\ex H:Prop; \ex K:H \ra Prop; \ex h_{3}:H;$\\
$(x:H)(K~x) = (x:h')(k'~x); (K~h_{3}) = B]$.\\
For $h_{2}$ we perform the substitution with head $v$ and $0$-splitting
$$h_{2} \la v$$
We get the context\\
$\Delta[\ex h':Prop; \ex k':h' \ra Prop; (I~((x:h')(k'~x))) = (I~(A \ra B));
\ex H:Prop; \ex K:H \ra Prop;$\\
$\ex h_{3}:H; (x:H)(K~x) = (x:h')(k'~x); (K~h_{3}) = B]$.\\
For $h'$ we perform the substitution with head $A$ and $0$-splitting
$$h' \la A$$
$\Delta[\ex k':A \ra Prop; (I~((x:A)(k'~x))) = (I~(A \ra B));
\ex H:Prop; \ex K:H \ra Prop; \ex h_{3}:H;$\\
$(x:H)(K~x) = (x:A)(k'~x); (K~h_{3}) = B]$.\\
For $k'$ we perform the substitution with head $B$ and $0$-splitting
$$k' \la [x:A]B$$
We get the context
$\Delta[\ex H:Prop; \ex K:H \ra Prop; \ex h_{3}:H;(x:H)(K~x) = (A \ra B); 
(K~h_{3}) = B]$.\\
For $H$ we perform the substitution with head $A$ and $0$-splitting
$$H \la A$$
We get the context
$\Delta[\ex K:A \ra Prop; \ex h_{3}:A;(x:A)(K~x) = (A \ra B); (K~h_{3}) = B]$.
\\
For $K$ we perform the substitution with head $B$ and $0$-splitting
$$K \la [x:A]B$$
We get the context
$\Delta[\ex h_{3}:A]$.\\
For $h_{3}$ we perform the substitution with head $w$ and $0$-splitting
$$h_{3} \la w$$
And we get the success context $\Delta$.\\
Let $\theta$ be the composition of all these substitutions, 
$\theta p = (u~(A \ra B)~v~w)$.

\section{Properties}

\subsection{Well-typedness}

In this section $\Gamma$ is a normal context well-formed in the $Meta$ type 
system such that the type the type of universal variables is $Prop$ or $Type$ 
but not 
$Extern$ and which is neither a success nor a failure context and 
$\sigma = x \la \gamma,t$ is a substitution of $\Sigma(\Gamma)$. 
We write $\Gamma = \Delta [\ex x:T] \Delta'$. We prove that the substitution 
$\sigma$ is well-typed in the context $\Gamma$. 

\begin{Prop} The context $\Delta \gamma$ is well-formed in the $Meta$ type 
system.

{\bd Proof} By induction on the length of $\gamma$, we check that the types 
of the existential variables and the constraints of $\gamma$ are well-typed in
the $Meta$ type system.

Since $(x_{1}:P_{1}) ... (x_{n}:P_{n})P$ is well-typed in the
$Meta$ type system, the terms $P_{i}$ are well-typed in the $Meta$ type system
of type $Prop$ or $Type$ but not $Extern$.
The term $(y_{1}:Q'_{1})... (y_{q}:Q'_{q})Q'$ is either the type of a 
universal
variable or one of the $P_{i}$, its type is therefore $Prop$ or $Type$ but not
$Extern$. So the types of the $Q'_{i}$ and of $Q'$ are $Prop$ or $Type$ but not
$Extern$. So the terms $Q_{i}$ and $Q$ are well-typed in the $Meta$ type
system.
The terms $(H_{i}~x_{1}~...~x_{n})$ and $(h~x_{1}~...~x_{n})$ are 
well-typed in the $Meta$ type system and have the type $s$ with 
$<s,s',s''> \in R$ for some $s'$ and $s''$. So $s$ is equal to $Prop$ or 
$Type$, but not to $Extern$.

So the terms 
$(\vec{x}:\vec{P})
(Q_{i}~(h_{1}~x_{1}~...~x_{n})~...~(h_{i-1}~x_{1}~...~x_{n}))$,
$(\vec{x}:\vec{P})s_{i}$,
$(\vec{x}:\vec{P})(z:(H_{i}~x_{1}~...~x_{n}))s'_{i}$,\\
$(\vec{x}:\vec{P})(Q~(h_{1}~x_{1}~...~x_{n})~...~(h_{q}~x_{1}~...~x_{n}))$,
$(\vec{x}:\vec{P})(K_{i-1}~x_{1}~...~x_{n}~(h_{q+i-1}~x_{1}~...~x_{n}))$,\\
$(\vec{x}:\vec{P})(z:(H_{i}~x_{1}~...~x_{n}))(K_{i}~x_{1}~...~x_{n}~z)$,
$(\vec{x}:\vec{P})(H_{i}~x_{1}~...~x_{n})$,\\
$(\vec{x}:\vec{P})(Q~(h_{1}~x_{1}~...~x_{n})~...~(h_{q}~x_{1}~...~x_{n}))$,
$(\vec{x}:\vec{P})(K_{r}~x_{1}~...~x_{n}~(h_{q+r}~x_{1}~...~x_{n}))$,
$(\vec{x}:\vec{P})P$,
$(\vec{x}:\vec{P})s$
and
$(\vec{x}:\vec{P})(y:(h~x_{1}~...~x_{n}))s'$ are well-typed in the 
$Meta$ type system and have the type $Prop$, $Type$ or $Extern$.
\end{Prop}

\begin{Prop} In the $Meta$ type system we have $\Delta \gamma \vdash t:T$.

{\bd Proof} 
For the substitution with $0$-splitting we have
$$[x_{1}:P_{1}] ... [x_{n}:P_{n}]
(w~(h_{1}~x_{1}~...~x_{n})~...~(h_{q}~x_{1}~...~x_{n})):
(\vec{x}:\vec{P})
(Q~(h_{1}~x_{1}~...~x_{n})~...~(h_{q}~x_{1}~...~x_{n}))$$
then using the constraint $\psi$
$$[x_{1}:P_{1}] ... [x_{n}:P_{n}]
(w~(h_{1}~x_{1}~...~x_{n})~...~(h_{q}~x_{1}~...~x_{n})):
(\vec{x}:\vec{P})P$$
For the substitution with $r$-splitting ($r \geq 1$), we prove by induction on
$i$ that in $\Delta \gamma$, for all $i \geq 1$
$$[x_{1}:P_{1}] ... [x_{n}:P_{n}]
(w~(h_{1}~x_{1}~...~x_{n})~...~(h_{q+i}~x_{1}~...~x_{n})):
(\vec{x}:\vec{P})(K_{i}~x_{1}~...~x_{n}~(h_{q+i}~x_{1}~...~x_{n}))$$
For $i = 1$ we have
$$[x_{1}:P_{1}] ... [x_{n}:P_{n}]
(w~(h_{1}~x_{1}~...~x_{n})~...~(h_{q}~x_{1}~...~x_{n})):
(\vec{x}:\vec{P})
(Q~(h_{1}~x_{1}~...~x_{n})~...~(h_{q}~x_{1}~...~x_{n}))$$
so using the constraint of $\chi_{1}$
$$[x_{1}:P_{1}] ... [x_{n}:P_{n}]
(w~(h_{1}~x_{1}~...~x_{n})~...~(h_{q}~x_{1}~...~x_{n})):
(\vec{x}:\vec{P})(z:(H_{1}~x_{1}~...~x_{n}))(K_{1}~x_{1}~...~x_{n}~z)$$
in the context $\Delta \gamma[\fa x_{1}:P_{1}; ... ;\fa x_{n}:P_{n}]$ we have
$$(w~(h_{1}~x_{1}~...~x_{n})~...~(h_{q}~x_{1}~...~x_{n})):
(z:(H_{1}~x_{1}~...~x_{n}))(K_{1}~x_{1}~...~x_{n}~z)$$
and since
$$h_{q+1}:(\vec{x}:\vec{P})(H_{1}~x_{1}~...~x_{n})$$
we deduce
$$(w~(h_{1}~x_{1}~...~x_{n})~...~(h_{q+1}~x_{1}~...~x_{n})):
(K_{1}~x_{1}~...~x_{n}~(h_{q+1}~x_{1}~...~x_{n}))$$
so in the context $\Delta \gamma$
$$[x_{1}:P_{1}] ... [x_{n}:P_{n}]
(w~(h_{1}~x_{1}~...~x_{n})~...~(h_{q+1}~x_{1}~...~x_{n})):
(\vec{x}:\vec{P})(K_{1}~x_{1}~...~x_{n}~(h_{q+1}~x_{1}~...~x_{n}))$$
Then if we assume this for $i$
$$[x_{1}:P_{1}] ... [x_{n}:P_{n}]
(w~(h_{1}~x_{1}~...~x_{n})~...~(h_{q+i}~x_{1}~...~x_{n})):
(\vec{x}:\vec{P})(K_{i}~x_{1}~...~x_{n}~(h_{q+i}~x_{1}~...~x_{n}))$$
using the constraint of $\chi_{i}$
$$[x_{1}:P_{1}] ... [x_{n}:P_{n}]
(w~(h_{1}~x_{1}~...~x_{n})~...~(h_{q+i}~x_{1}~...~x_{n})):
(\vec{x}:\vec{P})(z:(H_{i+1}~x_{1}~...~x_{n}))(K_{i+1}~x_{1}~...~x_{n}~z)$$
in the context $\Delta \gamma[\fa x_{1}:P_{1}; ... ;\fa x_{n}:P_{n}]$ we have
$$(w~(h_{1}~x_{1}~...~x_{n})~...~(h_{q+i}~x_{1}~...~x_{n})):
(z:(H_{i+1}~x_{1}~...~x_{n}))(K_{i+1}~x_{1}~...~x_{n}~z)$$
and since
$$h_{q+i+1}:(\vec{x}:\vec{P})(H_{i+1}~x_{1}~...~x_{n})$$
we deduce
$$(w~(h_{1}~x_{1}~...~x_{n})~...~(h_{q+i+1}~x_{1}~...~x_{n})):
(K_{i+1}~x_{1}~...~x_{n}~(h_{q+i+1}~x_{1}~...~x_{n}))$$
so in the context $\Delta \gamma$
$$[x_{1}:P_{1}] ... [x_{n}:P_{n}]
(w~(h_{1}~x_{1}~...~x_{n})~...~(h_{q+i+1}~x_{1}~...~x_{n})):
(\vec{x}:\vec{P})(K_{i+1}~x_{1}~...~x_{n}~(h_{q+i+1}~x_{1}~...~x_{n}))$$
So we deduce
$$[x_{1}:P_{1}] ... [x_{n}:P_{n}]
(w~(h_{1}~x_{1}~...~x_{n})~...~(h_{q+r}~x_{1}~...~x_{n})):
(\vec{x}:\vec{P})(K_{r}~x_{1}~...~x_{n}~(h_{q+r}~x_{1}~...~x_{n}))$$
then using the last constraint ($\psi$)
$$[x_{1}:P_{1}] ... [x_{n}:P_{n}]
(w~(h_{1}~x_{1}~...~x_{n})~...~(h_{q+r}~x_{1}~...~x_{n})):
(\vec{x}:\vec{P})P$$
In the same way, if $\sigma = x \la \gamma,t$ with
$t = [x_{1}:P_{1}]...[x_{n}:P_{n}]u$ where $u$ is a sort or a product then the
term $t$ also is well-typed and has also type $T$ in the context 
$\Delta \gamma$ in the $Meta$ type system.
\end{Prop}

\begin{Prop}
The substitution $\sigma$ is well-typed in the context $\Gamma$ in the $Meta$ 
type system.

{\bd Proof} Let us write $\Delta' = [e_{1}; ...; e_{n}]$.  
We prove by induction on $i$ that the substitution $\sigma$ is well-typed
in $\Delta [\ex x:T] [e_{1}; ...; e_{i}]$ in the $Meta$ type system.

We have $\sigma (\Delta [\ex x:T]) = \Delta \gamma$ and $\sigma T = T$, so 
the context $\sigma (\Delta [\ex x:T])$ is well-formed in the $Meta$ type 
system
and we have $\sigma (\Delta [\ex x:T]) \vdash t:\sigma T$. Obviously
$\sigma$ binds no universal variables of $\Delta [\ex x:T]$. So $\sigma$ is 
well-typed in $\Delta [\ex x:T]$. 

Let us assume now that the substitution $\sigma$ is well-typed in 
$\Delta [\ex x:T] [e_{1}; ...; e_{i}]$. The item $e_{i+1}$ is either the 
declaration of a variable $Qy:P$ or a constraint $a = b$. In the first case we
have
$$\Delta [\ex x:T] [e_{1}; ...; e_{i}] \vdash P:s$$
for some sort $s$, and in the second
$$\Delta [\ex x:T] [e_{1}; ...; e_{i}] \vdash a:U~~~\mbox{and}~~~
\Delta [\ex x:T] [e_{1}; ...; e_{i}] \vdash b:U$$
for some $U$.
So since the substitution $\sigma$ is well-typed in 
$\Delta [\ex x:T] [e_{1}; ...; e_{i}]$ we have in the first case
$$\sigma (\Delta [\ex x:T] [e_{1}; ...; e_{i}]) \vdash \sigma P:s$$
and in the second
$$\sigma(\Delta [\ex x:T] [e_{1}; ...; e_{i}]) \vdash \sigma a:\sigma U
~~~\mbox{and}~~~
\sigma(\Delta [\ex x:T] [e_{1}; ...; e_{i}]) \vdash \sigma b:\sigma U$$
So the context 
$\sigma (\Delta [\ex x:T] [e_{1}; ...; e_{i+1}]) =
\sigma (\Delta [\ex x:T] [e_{1}; ...; e_{i}]) [\sigma e_{i+1}]$
is well-formed. Then since we have $\Delta \gamma \vdash t:T$ and 
$\sigma$ obviously binds no universal variables of
$\Delta [\ex x:T] [e_{1}; ...; e_{i+1}]$, the substitution $\sigma$ is 
well-typed in $\Delta [\ex x:T] [e_{1}; ...; e_{i+1}]$ in the 
$Meta$ type system.
\end{Prop}

\subsection{Soundness}

\begin{Def} : Solution to a Context

Let $\Gamma$ be a well-formed context, the substitution $\theta$ is said to be 
a {\it solution} to $\Gamma$ if 
\begin{itemize}
\item 
the substitution $\theta$ is well-typed in $\Gamma$, 
\item
the context $\theta \Gamma$ is a success context.
\end{itemize}
\end{Def}

\begin{Def} Normal Solution to a Context

Let $\Gamma$ be a well-formed context, the substitution $\theta$, solution to 
$\Gamma$, is said to be a {\it normal solution} to $\Gamma$ if
\begin{itemize}
\item
the substitution $\theta$ binds exactly the existential variables of $\Gamma$,
\item for each existential variable $x$ of 
$\Gamma$, the context associated to $x$ by $\theta$ is empty and 
$\sigma x$ is normal $\eta$-long in the context $\theta \Delta$ where 
$\Delta$ is the unique context such that 
$\Gamma = \Delta [\ex x:T] \Delta'$.
\end{itemize}
\end{Def}

\begin{Def} Normal Form of a Solution

Let $\Gamma$ be a well-formed context and $\theta$ a solution to $\Gamma$.
We define $\theta'$ the {\it normal form of $\theta$ in $\Gamma$} by induction
over the length of $\Gamma$.
\begin{itemize}
\item
If $\Gamma = [~]$ then we let $\theta' = \{\}$.
\item
If $\Gamma = \Delta [\fa x:T]$ or $\Gamma = \Delta [a = b]$ then 
$\theta$ is a solution to $\Delta$, we let $\theta'$ be the normal form
of $\theta$ in $\Delta$.
\item
If $\Gamma = \Delta [\ex x:T]$ then $\theta$ is a solution to $\Delta$. We
let $\theta_{1}$ be the normal form of $\theta$ in $\Delta$. 
We have $\theta \Gamma \vdash \theta x:\theta T$, let $t$ be the normal
$\eta$-long form of $\theta x$ in $\theta \Gamma$. We let 
$\theta' = \theta_{1} \cup \{<x,[~],t>\}$.
\end{itemize}
\end{Def}

\begin{Prop}
Let $\Gamma$ be a context and $\Gamma'$ be a context obtained by removing 
some constraints relating identical terms.
If  $\Gamma$ is well-formed then $\Gamma'$ is well-formed. If 
$\Gamma \vdash t:T$ then $\Gamma' \vdash t:T$.

{\bd Proof} By induction on the length of the derivation of 
{\it $\Gamma$ well-formed} or $\Gamma \vdash t:T$.
\end{Prop}

\begin{Prop}
Let $\Delta$ and $\gamma$ be two contexts such that $\Delta \gamma$ is a 
success context and $\gamma$ is an existential context. Then $\gamma$ is a 
list of constraints relating terms well-typed without using the constraints and
whose normal forms are identical. 

{\bd Proof} By induction on the length of $\gamma$.
\end{Prop}

\begin{Prop} Let $\Gamma$ be a well-formed context and $\theta$ a solution to 
$\Gamma$. Let $\theta'$ be the normal form of $\theta$. Then $\theta'$ is a
normal solution to $\theta$.

{\bd Proof} By induction on the length of $\Gamma$.
\end{Prop}

\begin{Lem} 
Let $\Gamma$ be a context, if there exists a derivation 
$[\sigma_{1}; ...; \sigma_{m}]$ of $\Gamma$, then there exists a normal 
solution to $\Gamma$. 

{\bd Proof}
The substitution $\sigma_{m} \circ ... \circ \sigma_{1}$, is a solution to
$\Gamma$. Let $\theta$ be its normal form.
\end{Lem}

The substitution $\theta$ called the substitution {\it denoted} by 
the derivation $[\sigma_{1}; ...; \sigma_{m}]$.

\medskip

Now we prove that the substitution $\theta$ which is well-typed in the $Meta$
type system is also well-typed in the original system ${\cal T}$.

\begin{Prop}
Let $\Gamma$ be a context well-formed in a type system ${\cal T}$ 
and $T$ be either a term well-typed in $\Gamma$ in the system ${\cal T}$
or the symbol $Type$. 
Let $t$ be a normal $\eta$-long term such that $\Gamma \vdash t:T$ in the 
$Meta$ type system such that for every subterm of $t$ which is a product 
$(x:U)U'$ 
if we let $s$ be the type of $U$, $s'$ be the type of $U'$ and $s''$ be the
type of $(x:U)U'$, we have $<s,s',s''> \in R$ (the set of rules of the
system ${\cal T}$) and for every subterm of $t$ which is a sort $s$ we have 
$s = Prop$ (and not $Type$), then
we have $\Gamma \vdash t:T$ in the system ${\cal T}$.

{\bd Proof} By induction over the structure of $t$.
\begin{itemize}
\item
If $t = [x:U]u'$ then $T = (x:U)U'$ is well-typed in the type system
${\cal T}$ so $U'$ is well-typed in $\Gamma [\fa x:U]$ in the
type system ${\cal T}$ and by induction hypothesis 
$\Gamma [\fa x:U] \vdash u':U'$ in the system ${\cal T}$. 
So $\Gamma \vdash t:T$ in the system ${\cal T}$.
\item
If $t = (x:U)U'$ then by induction hypothesis $U$ and $U'$ are
well-typed in the system ${\cal T}$ and since the rule $<s,s',s''>$ is a
rule of the system ${\cal T}$, we have $\Gamma \vdash t:T$ in the system
${\cal T}$.
\item
If $t = (x~c_{1}~...~c_{n})$ with $x$ variable declared in $\Gamma$
then $x$ is well-typed in $\Gamma$ in the system ${\cal T}$ and 
we prove by induction on $i$ that the type of $c_{i}$ is well-typed in
the system ${\cal T}$ and $c_{i}$ is well-typed in the system ${\cal T}$
and we conclude that $\Gamma \vdash t:T$ in the system ${\cal T}$.
\item
If $t$ is a sort then by hypothesis $t = Prop$, it is well-typed in the 
system ${\cal T}$.
\end{itemize}
\end{Prop}

\begin{Prop}
Let $\Gamma$ be a well-formed context in the system
${\cal T}$, $[\sigma_{1}; ...; \sigma_{m}]$ be a derivation of $\Gamma$
and $\theta$ the substitution denoted by this derivation.
Let $x$ be an existential variable of $\Gamma$ and $t = \theta x$.
For every subterm of $t$ which is a product $(x:U)U'$
if we let $s$ be the type of $U$, $s'$ be the type of $U'$ and $s''$ be the
type of $(x:U)U'$, we have $<s,s',s''> \in R$ (the set of rules of the
system ${\cal T}$) and for every subterm of $t$ which is a sort $s$ we have
$s = Prop$ (and not $Type$),

{\bd Proof} By induction on $m$.
\end{Prop}

\begin{Prop} Let $\Gamma$ be a context well-formed in the system 
${\cal T}$, $[\sigma_{1}; ...; \sigma_{m}]$ a derivation of $\Gamma$
and $\theta$ the substitution denoted by this derivation. 
The substitution $\theta$ is well-typed in $\Gamma$ in the system ${\cal T}$.

{\bd Proof} By induction on the length of $\Gamma$.
\end{Prop}

\begin{Theo} Soundness

Let $\Gamma$ be a (non constrained, non quantified) context and $P$ a 
well-typed type in $\Gamma$.
Let $\Gamma' = \Gamma[\ex x:P]$. If there exists a derivation of
$\Gamma'$ then there exists a proof $t$ of $P$ in $\Gamma$ in ${\cal T}$.

{\bd Proof} Let $\theta$ be the substitution denoted by the derivation of 
$\Gamma'$. The substitution $\theta$ is well-typed in $\Gamma'$
in the system ${\cal T}$ and is a normal solution to $\Gamma'$.
Let $t = \theta x$. 
Since $\theta$ is normal we have $\theta \Gamma' = \theta \Gamma$.
We have $\theta \Gamma' = \theta \Gamma$, $\theta \Gamma = \Gamma$,
$\theta P = P$ and $\theta \Gamma'  \vdash \theta x:\theta P$
so $\Gamma \vdash t:P$.
\end{Theo}

The proof $t$ is called the proof {\it denoted} by the derivation.

\subsection{Completeness}

\begin{Def} The Relation $<$

Let $<$ be the smallest transitive relation defined on normal $\eta$-long 
terms such that
\begin{itemize}
\item
if $t_{\Gamma}$ is a strict subterm of $t'_{\Delta}$ then 
$t_{\Gamma} <t'_{\Delta}$,
\item
if $T_{\Gamma}$ is the normal $\eta$-long form of the type of 
$t_{\Gamma}$ in $\Gamma$ and $t$ is not a sort then $T_{\Gamma} < t_{\Gamma}$.
\end{itemize}
\end{Def}

As proved in \cite{matching} \cite{These}, this relation is well-founded, 
which means 
that a function that recurses both on a strict subterm and on the type of its
argument is total.

\begin{Def} Size of a Term

Let $\Gamma$ be a context and $t$ a term well-typed in $\Gamma$.
Let $T$ be the normal $\eta$-long form of the type of $t$ in $\Gamma$.
We define by induction over $<$, the {\it size} of $t_{\Gamma}$ 
($|t_{\Gamma}|$)
\begin{itemize}
\item
If $t$ is a sort then $|t_{\Gamma}| = 1$.
\item
If $t$ is a variable then $|t_{\Gamma}| = |T_{\Gamma}|$.
\item
If $t = (u~v)$ then 
$|t_{\Gamma}| = |u_{\Gamma}| + |v_{\Gamma}| + |T_{\Gamma}|$.
\item
If $t = [x:U]u$ then $|t_{\Gamma}| = |u_{\Gamma[x:U]}|$.
\item
If $t = (x:U)V$ then 
$|t_{\Gamma}| = |U_{\Gamma}| + |V_{\Gamma[x:U]}| + |T_{\Gamma}|$.
\end{itemize}
\end{Def}

\begin{Def} Size of a Substitution 

The {\it size} of a substitution 
$\theta = \{<x_{i},\gamma_{i},t_{i}>\}$ is the sum of the sizes of the $t_{i}$.
\end{Def}

\begin{Prop} If $\Gamma$ is a failure context, then for every 
substitution $\sigma$, $\sigma \Gamma$ is also a failure context
and therefore is not a success context.

{\bd Proof} Let $a = b$ be a constraint of $\Gamma$ relating two
ground different terms, then $(\sigma a = \sigma b)$ is $(a = b)$ which is a
constraint relating two well-typed ground different terms.
\end{Prop}

\begin{Lem}
Let $\Gamma$ be a well-formed context and $\theta$ a normal solution to 
$\Gamma$, then there exists a derivation of $\Gamma$ which denotes $\theta$.

{\bd Proof} By induction on the size of $\theta$.

The context $\Gamma$ is not a failure context because there exists a 
substitution $\theta$ such that $\theta \Gamma$ is a success context.
If it is a success context then $[~]$ is a derivation of $\Gamma$.
Otherwise let us chose an existential variable $x$ with a normal $\eta$-long
type $(x_{1}:P_{1}) ... (x_{n}:P_{n})P$ such that $P$ is atomic rigid.
Let $t = \theta x$. Since $\theta$ is well-typed in $\Gamma$,
$\theta \Gamma \vdash t:(x_{1}:\theta P_{1}) ... (x_{n}:\theta P_{n})\theta P$.
Since the term $P$ is atomic rigid, the term $\theta P$ is atomic.
Then $t = [x_{1}:\theta P_{1}] ... [x_{n}:\theta P_{n}]u$ 
with $u$ atomic or a product ($\eta$-long form).
\begin{itemize}
\item
If $u = (w~u_{1}~...~u_{p})$ then let $q$ the number of products
of the type of $w$ and $r = p-q$.
Since the type of $u$ is atomic we have $p \geq q$, i.e. $r \geq 0$.
We let
$$\sigma = \{<x,\gamma,[x_{1}:P_{1}] ... [x_{n}:P_{n}]
(w~(h_{1}~x_{1}~...~x_{n})~...~(h_{p}~x_{1}~...~x_{n}))>\}$$
with $\gamma$ defined in the algorithm.
Then we build the terms to be substituted to the
variables $h_{i}$ ($1 \leq i \leq p$) and $H_{i}$ and $K_{i}$ 
($1 \leq i \leq r$). We let
$$u'_{i} = [x_{1}:P_{1}] ... [x_{n}:P_{n}]u_{i}$$
And for all $1 \leq i \leq r$, $(w~u_{1}~...~u_{q+i-1})$ has a type which is a
product, let $(y:U_{i})V_{i}$ be this product.
We let
$$U'_{i} = [x_{1}:P_{1}] ... [x_{n}:P_{n}]U_{i}$$
$$V'_{i} = [x_{1}:P_{1}] ... [x_{n}:P_{n}][y:U_{i}]V_{i}$$
$$\theta' = \theta - \{<x,[~],\theta x>\} \cup \{<h_{i},[~],u'_{i}>\}
\cup \{<H_{i},[~],U'_{i}>,<K_{i},[~],V'_{i}>\}$$
We have $\theta = \theta' \circ \sigma$.
Let us prove that the substitution $\theta'$ is smaller than $\theta$.
We have 
$$\theta x = [x_{1}:\theta P_{1}] ... [x_{n}:\theta P_{n}](w~u_{1}~...~u_{p})$$
Let $T_{i}$ be the type of $(w~u_{1}~...~u_{i})$.\\
$|\theta x| = |(w~u_{1}~...~u_{p})|
= |T_{0}| + |u_{1}| + |T_{1}| + ... + |u_{p}| + |T_{p}|$\\
$> |T_{q}| + ... + |T_{p-1}| + |u_{1}| + ... + |u_{p}|
= |(y:U_{1})V_{1}| + ... + |(y:U_{r})V_{r}| + |u_{1}| + ... + |u_{p}|$\\
$> |U_{1}| + ... + |U_{r}| + ... + |V_{1}| + ... + |V_{r}| + |u_{1}| + 
... + |u_{p}|$\\
$= |U'_{1}| + ... + |U'_{r}| + |V'_{1}| + ... + |V'_{r}| + |u'_{1}| +
... + |u'_{p}|$.\\
So $|\theta| > |\theta'|$.
\item
If $u = (x:U)V$ then we let
$$\sigma = \{<x,\gamma,[x_{1}:P_{1}] ... [x_{n}:P_{n}]
(y:(h~x_{1}~...~x_{n}))(k~x_{1}~...~x_{n}~y)>\}$$
with $\gamma$ defined in the algorithm.
Then we build the terms to be substituted to the variables $h$ and $k$. 
We let
$$U' = [x_{1}:P_{1}] ... [x_{n}:P_{n}]U$$
$$V' = [x_{1}:P_{1}] ... [x_{n}:P_{n}][y:U]V$$
$$\theta' = \theta - \{<x,[~],\theta x>\} \cup \{<h,[~],U'>,<k,[~],V'>\}$$
We have $\theta = \theta' \circ \sigma$.
Let us prove that the substitution $\theta'$ is smaller than $\theta$.
We have 
$$\theta x = [x_{1}:\theta P_{1}] ... [x_{n}:\theta P_{n}](y:U)V$$
$|\theta x| = |(y:U)V| > |U| + |V| = |U'| + |V'|$.\\
So $|\theta| > |\theta'|$.
\end{itemize}
In both cases, by induction hypothesis, there exists a 
derivation $D$ of $\sigma \Gamma$ that denotes $\theta'$ 
and $[\sigma]D$ is a derivation of $\Gamma$ that denotes $\theta$.
\end{Lem}

\begin{Theo} Completeness

Let $\Gamma$ be a (non constrained, non quantified) context
and $P$ a type well-typed in $\Gamma$ 
such that there exists a term $t$ such that $\Gamma \vdash t:P$ then there
exists a derivation of $\Gamma' = \Gamma[\ex x:P]$ which denotes the normal
$\eta$-long form of the proof $t$.

{\bd Proof} Let $t'$ be the normal $\eta$-long form of $t$, and 
$\theta = \{<x,[~],t'>\}$. The substitution 
$\theta$ is a normal solution to $\Gamma'$, so there exists a derivation of 
$\Gamma'$ that denotes $\theta$.
\end{Theo}

\section{Efficiency Improvements}

To improve the efficiency of this method we have to recognize nodes that
cannot lead to success nodes and prune the search tree.

\subsection{Incremental Partial Checking of Constraints}

First we must perform incremental partial checking of the constraints.
So we define a function of simplification, which is very similar to the
SIMPL function from \cite{Huet75} \cite{Huet76}.
In order to define this function we have to modify a little the syntax of
constraints: a constraint is now a triple $<\delta,a,b>$ where $\delta$ is a
context with universal variables only. If 
$\Gamma = \Delta[<\delta,a,b>]\Delta'$ then the terms $a$ and $b$ must be
well-typed and have the same type in $\Delta \delta$.

\begin{Def} Simplification

We consider only constraints well-typed without the constraints.
While there are rigid-rigid constraints in $\Gamma$ and the simplification 
has not failed we iterate the process of replacing $\Gamma$ by $\Gamma'$.

Let $\delta,a = b$ be a rigid-rigid constraint of $\Gamma$.
$\Gamma = \Delta[\delta,a = b]\Delta'$.
We define $\Gamma'$ as
\begin{itemize}
\item
if $a$ and $b$ are both abstractions $a = [x:T]a'$, $b = [x:U]b'$,
since the terms $a$ and $b$ have the same type in $\Delta \delta$ we have
$T = U$,
we let $\Gamma' = \Delta[\delta [\fa x:T], a' = b']\Delta'$,
\item
if $a$ and $b$ are both products $a = (x:T)a'$, $b = (x:U)b'$ and $T$ 
and $U$ have the same type,
we let $\Gamma' = \Delta[\delta,T = U][\delta [\fa x:T], a' = b']\Delta'$,
\item
if $a$ and $b$ are both atomic with the same head variable and they
have the same number of arguments, 
$a = (w~u_{1}~...~u_{p})$, $b = (w~t_{1}~...~t_{p})$,
we let 
$\Gamma' = \Delta[\delta, u_{1} = t_{1}]...[\delta, u_{p} = t_{p}]\Delta'$,
\item
in the other cases the simplification fails.
\end{itemize}
If the simplification fails then the branch is hopeless, it can be pruned.
\end{Def}

\subsection{Using the Flexible-Rigid Constraints}

When we have a well-typed flexible-rigid constraint, and we can solve the head 
variable of
the flexible term (i.e. when the head of its type is rigid) we must solve it 
first, the only candidates for the head variable in an elementary substitution
are the bound variables and the head of the rigid term \cite{Huet75} 
\cite{Huet76}.  The other variables lead to unsolvable rigid-rigid constraints.

\subsection{Using the Flexible-Flexible Constraints}

When we have a well-typed flexible-flexible constraint and we can solve one of 
the head variables, we must solve it first, the constraint does not help to 
restrict the number of substitutions, but it may become a flexible-rigid
constraint the next step. Moreover solving constraints helps to unfreeze types 
of existential variables and constraints that are not well-typed without the
constraints.

\subsection{Avoiding Splitting}

The term $(Q~(h_{1}~x_{1}~...~x_{n}) ... (h_{q}~x_{1}~...~x_{n}))$ in the 
definition of the method is atomic, if it is rigid then all the solutions with
$r$-splitting $r \neq 0$ lead
to hopless constraints and therefore they can be avoided.
In particular in the $\lambda \Pi$-calculus, we never consider 
existential variables for types or predicates and so splitting can always be
avoided.

\subsection{Solving some Decidable Unification Problems}

When we have a constraint well-typed without the constraints which is either
a first order unification problem \cite{Robinson},
an argument-restricted unification problem \cite{Miller} \cite{Pfenning91},
a second order matching problem \cite{Huet76} \cite{HueLan} \cite{matching}
\cite{These},
a second-order-argument-restricted matching problem \cite{matching} 
\cite{These} then we can solve it using always terminating algorithms
and apply the substitutions obtained in this way. 
In a system with such heuristics, the elimination of a first-order universal
quantifier is a one-step operation and the elimination of a higher-order 
universal quantifier is a more complex operation that needs several steps and 
can lead to backtracking.

\subsection{Priority to the Rightmost Variable}

When we have two existential variables $x$ and $y$ such that $y$ is declared 
on the right of $x$ and $x$ has an occurrence in the type of $y$, if both
variables may be instantiated then we have to begin with the rightmost.
For instance if we have an axiom
$u:(P~n)$ and we search an integer $x$ such that $(P~x)$. 
We have two existential variables: $x:Nat$ and $y:(P~x)$.
If we instantiate $y$ by $u$ then $x$ will be automatically instantiated by
$n$. But if we begin by instantiating $x$, we will instantiate $x$ by
$0$, $1$, $2$, etc. and fail to prove $(P~0)$, $(P~1)$, $(P~2)$, etc. before
we reach $x = n$.

\subsection{Normalizing some ill-typed terms}

An important improvement of the method would be to recognize some normalizable 
ill-typed terms and then be able to use the constraints more efficiently. 
For instance it is possible to prove that if $t$ is a normal $\eta$-long term 
and $\sigma$ an elementary  substitution 
$x \la [x_{1}:P_{1}] ... [x_{n}:P_{n}]u$ where $u$ 
is a product or an atomic 
term with a head variable which is a universal variable or a sort (but not
one of the $x_{i}$), then $\sigma t$ is normalizable.

\section{Extension to The Calculus of Constructions With Universes}

\begin{Def} The Calculus of Constructions With Universes and Without
Cumulativity

The {\it Calculus of Constructions With Universes and Without Cumulativity}
is a pure type system which has an infinite number of sorts : 
$Prop, Type_{0} (= Type), Type_{1}, Type_{2} , Type_{3}, ...$, the axioms 
$Prop:Type_{0}$ and $Type_{i}:Type_{i+1}$ and the rules 
$$<Prop,Prop,Prop>, <Type_{i},Prop,Prop>,$$
$$<Prop,Type_{i},Type_{i}>, <Type_{i},Type_{j},Type_{max\{i,j\}}>$$
\end{Def}

The strong normalization and confluence of $\beta \eta$-reduction for 
this system are not yet fully proved. But since all the terms typable in this
system are typable in the Calculus of Constructions with Universes 
\cite{Coquand86} \cite{Luo}, the $\beta$-reduction is strongly normalizable 
and confluent \cite{Coquand86} \cite{Luo}.
It is conjectured in \cite{Geuvers} that in a pure type system in which the
$\beta$-reduction is strongly normalizable the 
$\beta \eta$-reduction also is. Then assuming strong normalization, 
confluence is proved in \cite{Geuvers} \cite{Salvesen} \cite{Coquand91}.

In an extension of the method presented here to the Calculus of Constructions 
With Universes and Without Cumulativity we do not need a $Meta$ type system any
more because all the terms we consider in the algorithm in this system are 
well-typed in this system. 
More generaly the forumulation of this method for an arbitrary normalizable
pure type system
${\cal T}$ seems to requires only the definition of a $Meta$ type system for 
${\cal T}$.

In a system which has a large number of sorts, there may be a
large number of substitutions of the form $\sigma = x \la \gamma,t$ with
$t = [x_{1}:P_{1}]...[x_{n}:P_{n}]u$ where $u$ is a sort or a product.
For instance, in the Calculus of Constructions With Universes
and Without Cumulativity, we can instantiate a variable $x:Prop$ by an
infinite number of terms $(y:h)(k~y)$ with $h:Type(i)$ and $k:(y:h)Prop$. 
In order to avoid more inefficiencies it seems possible to use sort variables
and constraints on these variables as in the method of floating universes
\cite{HarPol} \cite{Huet88}.

\section{An Incomplete but more Efficient Method}

In the example in section 5.2, in order to get the proof 
$(u~(A \ra B)~v~w):B$ 
we have considered the elementary substitution with $1$-splitting
$$p \la (u~h_{1}~h_{2}~h_{3})$$
with three applications although the type of $u$ begins by only two products.

Because we have to consider all these possible degrees of splitting,
the method is quite inefficient,
we get a much more efficient one if we restrict the rule to elementary 
substitutions with $0$-splitting and a number of abstractions ranging from 
$0$ to the number of products of the type of the existential variable we 
substitute. 
Let us call this algorithm {\it algorithm 
with weak splitting} (by opposition the previous one can be called {\it 
algorithm with strong splitting}). The search tree of the weak splitting 
algorithm is finitely branching.

This algorithm with weak splitting cannot synthesize a proof of $B$, 
but it can synthesize a proof of $A \ra B$.
Indeed with the variable $p':A \ra B$ we perform the substitution
$$p' \la (u~h_{1}~h_{2})$$
with the types $h_{1}:Prop$, $h_{2}:(I~h_{1})$
(remark that we have no abstraction although the type of $p'$ 
begins by a product). The constraint $h_{1} = A \ra B$ suggests 
the substitution
$$h_{1} \la (A \ra B)$$
then we have an existential variable $h_{2}:(I~(A \ra B))$ we perform the
substitution
$$h_{2} \la v$$
The synthesized proof is $(u~(A \ra B)~v):(A \ra B)$. Remark that this term is 
not in $\eta$-long form, this form is $[w:A](u~(A \ra B)~v~w):(A \ra B)$.

In both proofs considered above (of $B$ and $A \ra B$) the problem is to 
remark that the 
variable $u$ of type $(P:Prop)((I~P) \ra P)$ must be applied to the proposition
$(A \ra B)$.
With the strong splitting algorithm this is found in both cases, but the price
to pay is inefficiency, with the weak splitting this is found when the type
of the existential variable is $A \ra B$ but not when it is $B$.
Roughly speaking, this means that when we must use a proposition with a 
quantifier on a 
proposition or a predicate (for instance an induction axiom) the weak 
splitting algorithm finds the proposition to be used only when it
{\it can be seen} in the type of the existential variable, 
and not when the proof requires an {\it induction 
loading} \cite{Helmink} (as here from $B$ to $A \ra B$).

This weak splitting algorithm is quite similar to the introduction-resolution
method \cite{Helmink} for some unification algorithm. 
Actually some proofs that cannot be synthesized by introduction-resolution can 
be synthesized with weak splitting. For instance in a context that contains 
the universal variables $u:(P:Prop)(P \ra A)$ and $v:(B \ra C)$ the proof
$(u~(B \ra C)~[x:B](v~x)):A$ is synthesized by the weak-splitting algorithm
and not by the introduction-resolution method. To get exactly the 
introduction-resolution method we would have to forbid completely product 
substitutions.

This method with weak splitting is incomplete but its transitive closure is
complete \cite{These}. This means that even if it cannot be synthesized, each
proof can be broken up in smaller proofs that can be synthesized. 

\section{Unification}

\subsection{Ground Unification}

The unification problem is to decide if given a context $\Gamma$ 
(with universal and existential variables but no constraints)
and two terms $a$ and $b$ well-typed in $\Gamma$ with the same type, 
there exists a substitution $\theta$ well-typed in 
$\Gamma$, such that for every variable $x$ bound by $\theta$, the context 
associated to $x$ by $\theta$ contains no constraints and such that 
$\theta a = \theta b$. In first order unification and 
higher order unification we do not require $\theta$ to fill all the 
existential variables of $\Gamma$, moreover the substitution $\theta$ may 
introduce new existential variables.
This tolerance is due to the fact that in first order and higher order logic, 
types are all supposed to be inhabited, so from any unifier we can deduce a 
ground unifier by filling the unfilled variables with arbitrary terms. So 
unifiability is equivalent to ground unifiability.
This is not the case any more in type systems where we have empty types. 
In unification in simply typed $\lambda$-calculus under a quantified context
\cite{Miller} types may also be empty and unifiability is not equivalent 
to ground unifiability. 
The notion of unifiability considered in \cite{Miller} is the one of ground 
unifiability.

A semi-algorithm for deciding ground unifiability is to apply the method 
developed in this paper to the context $\Gamma [a = b]$.
Moreover this algorithm enumerates all the ground unifiers. 

In contrast with what happens in simply typed $\lambda$-calculus, we have to 
solve flexible-flexible equations. Indeed these equations may have no solution
since the head variables of the terms may have empty types. Moreover it is 
proved in \cite{Miller} that the problem of deciding the existence of 
solutions for
flexible-flexible unification problems under a quantified context is 
undecidable in simply typed $\lambda$-calculus. This proof generalizes easily
to type systems.

\subsection{Toward Open Unification}

The utility of an open unification algorithm is not obvious, since in a
proof-search algorithm (or more generally in an algorithm that 
uses unification) existential variables are introduced to be filled-in and 
not to remain forever.

If we still need such an algorithm we may remark that flexible-flexible 
equations have to be solved since, in contrast with what happens in 
simply typed $\lambda$-calculus, they may have no solution.
Consider
$$[\fa A:Prop; \fa B:Prop; \fa u:A; \ex x:(P:Prop)P]$$
$$a = (x~(A \ra B)~u)~~~~~b = (x~B)$$
The equation $a = b$ is flexible-flexible, but there is no solution to this 
problem. Indeed, let us suppose there is one and consider $t = \theta x$. 
We have $t:(P:Prop)P$ so $t = [P:Prop]t'$ with $t':P$. So $t'$ is neither an 
abstraction nor a product. It is an atomic term $(f~c_{1}~...~c_{n})$. 
If $f$ were the variable $P$ we would have $n = 0$ and thus $t' = P$ which 
is impossible for type reasons, so $f \neq P$.
Since $(t'[P \la (A \ra B)]~u) = t'[P \la B]$ we have
$$(f~c_{1}[P \la (A \ra B)]~...~c_{n}[P \la (A \ra B)]~u) = 
(f~c_{1}[P \la B]~...~c_{n}[P \la B])$$
So $(f~c_{1}[P \la (A \ra B)]) = f$ which is impossible since a variable 
cannot be equal to an application.

Searching for solutions of flexible-flexible equations seems to be rather 
difficult,
since the head variable of an elementary substitution may now be 
any variable of $\Gamma$ and also any new variable.
We would have to design a proof search algorithm similar to the one described 
in this paper but that enumerates all the proofs of a proposition, under all
the possible sets of axioms.

It may be possible to recognize flexible-flexible equations that have 
solutions and those that do not have any and keep the former unsolved until the
end of the process as we do in unification in simply typed $\lambda$-calculus.
In this case the normalization of terms well-typed using such constraints is 
not obvious, we may have to perform one normalization step as an elementary 
operation of the algorithm.
Also some product substitutions have to be performed when the types 
of the variables to be instantiated in flexible-rigid and flexible-flexible
equations has an existential head variable.

In contrast with ground unifiability, open unifiability seems a quite
difficult problem with small interest.

\section*{Conclusion}

In this paper we have described an semi-decision procedure for type systems.
In fact, since proof checking is decidable in these systems, the existence of
such a procedure is obvious: we just need to enumerate all the lists of 
characters until we get a proof of the proposition to be proved.
Of course this method (which is sometime used to prove the semi-decidability 
of predicate calculus) is of no practical interest.
We can let it be more realist in enumerating not all the lists of characters
but only the normal $\lambda$-terms (or proof-trees), this method is comparable
to the methods of the early sixties based on the search of a counter Herbrand
model. 

Resolution excludes more hopeless tentatives by remarking that in a
premise $\fa x.(P~x)$, we need to instantiate $x$ by $t$ only if $(P~t)$ is 
an instance of the proposition to be proved or an hypothesis of another 
premise.
This method formulated by Robinson \cite{Robinson} for first order logic and 
by Huet \cite{Huet72} for higher order logic is generalized here to type 
systems. The notions of proof-term that appear in these systems simplify
the method and makes more explicit the idea of blind enumeration of proof-terms
regulated by a failure anticipation mechanism exploiting type constraints.

\section*{Acknowledgements}

The author thanks G\'erard Huet who has supervised this work and
Amy Felty, Herman Geuvers, Christine Paulin and the anonymous referees for
many helpful comments and criticisms.

\end{document}